\documentclass[referee]{aa} 
\usepackage{natbib}
\usepackage{graphicx}
\usepackage{txfonts}
\usepackage{url}

\usepackage[T1]{fontenc}

\begin{document}

\title{Stellar populations in a standard ISOGAL field in the Galactic disk\thanks{Based on observations with ISO, an ESA project with instruments funded by ESA Member States (especially the PI countries: France, Germany, the Netherlands and the United Kingdom) and with the participation of ISAS and NASA.}}

\author{S. Ganesh\inst{1,2}
\and A. Omont\inst{1}
\and U.C. Joshi\inst{2}
\and K.S. Baliyan\inst{2}
\and M. Schultheis\inst{3,1}
\and F. Schuller\inst{4,1}
\and G. Simon\inst{5}
}

\offprints{S. Ganesh, shashi@iap.fr, shashi@prl.res.in}

\institute{
Institut d'Astrophysique de Paris, CNRS and Universite Paris 6, 98bis boulevard Arago,
F-75014, Paris, France
\and Physical Research Laboratory, Navrangpura, Ahmedabad-380 009, India
\and Observatoire de Besan\c{c}on, Besan\c{c}on, France
\and Max-Planck-Institut fur Radioastronomie, Auf dem Hugel 69, D-53121 Bonn
\and GEPI, UMS-CNRS 2201, Observatoire de Paris, France
}

\date{Received date / Accepted date }

\titlerunning{A standard ISOGAL field}
\authorrunning{S. Ganesh et al.}

\abstract
   {} 
   {We aim to identify the stellar populations (mostly red giants and young stars) detected in the ISOGAL survey at 7 and 15$\mu$m towards a field (LN45) in the direction $\ell=-45, b=0.0$.}
   {The sources detected in the survey of the Galactic plane by the Infrared Space Observatory are characterized based on colour-colour and colour-magnitude diagrams.  We combine the ISOGAL catalog with the data from surveys such as 2MASS and GLIMPSE.  Interstellar extinction and distance are estimated using the red clump  stars detected by 2MASS in combination with the isochrones for the AGB/RGB branch.  Absolute magnitudes are thus derived and the stellar populations are identified based on their absolute magnitudes and their infrared excess.  }
   {A standard approach to the analysis of ISOGAL disk observations has been established.    We identify several hundred RGB/AGB stars and 22 candidate young stellar objects in the direction of this field in an area of 0.16 deg$^2$.  An over-density of stellar sources is found at distances corresponding to the distance of the Scutum-Crux spiral arm.  In addition, we determine mass-loss rates of AGB-stars using dust radiative transfer models from the literature.}
   {}
\maketitle


\section{Introduction}

	ISOGAL survey data at 7 and 15~$\mu$m of about 16deg$^2$ in selected fields of the inner
Galactic disk have allowed the detection of about 1~$\times$~10$^5$ point sources down to $\sim$10~mJy at
15~$\mu$m and 7~$\mu$m. In conjunction
with ground based near-IR surveys (DENIS, 2MASS), they offer the possibility to
investigate the different populations of infrared stars in the Galactic disk up to 15~$\mu$m. The
technical characteristics of the five wavelength ISOGAL-DENIS point source catalogue are presented
and discussed in detail by \citet{Schuller2003}, while \citet{Omont2003} have reviewed the
scientific capabilities and the main outcome of ISOGAL.

	The best detected stellar class is that of AGB stars which are almost completely
detected above the RGB tip at least at 7~$\mu$m up to the Galactic Center \citep{Glass1999,Omont1999}. The
infrared color ($K_S$-[15])$_\mathrm{0}$ is a very good measure of their mass-loss rate \citep{Ojha2003}. The mass-loss rate can also be derived from the pure ISOGAL color [7]-[15], which is practically independent of
extinction. Combined with variability data from MACHO\citep{Alcock1997} or EROS\citep{Palanque1998}, ISOGAL data of bulge fields have
shown that practically all sources detected at 15~$\mu$m are long period variables with interesting correlations between the mass-loss rate traced by the 15~$\mu$m excess and variability intensity and period
\citep{Alard2001}. The 15~$\mu$m data from the fields observed (0.29 deg$^2$ in total) in the Galactic bulge
($\vert b \vert \geq 1^\circ$) have been used by \citet{Ojha2003} to infer
global properties of the mass returned to the interstellar medium by AGB stars in the bulge. However,
similar work has not yet been carried out in the disk ISOGAL fields because of the difficulty in
properly estimating distances along the line of sight.

Red giants are by far the most numerous population of luminous bright stars in the
near and mid-infrared.  They are detectable through out the Galaxy with modern surveys such as 2MASS, DENIS, GLIMPSE,
ISOGAL and MSX. The reddening of various infrared colours may be used to trace the interstellar
extinction, A$_\mathrm{V}$. However, the near-IR colours $J-K_S$ or $H-K_S$ are generally the most useful because of
greater sensitivity to extinction with A$_\mathrm{J}$-A$_\mathrm{Ks}$~$\sim$~0.17\,A$_\mathrm{V}$ and
A$_\mathrm{H}$-A$_\mathrm{Ks}$~$\sim$~0.06\,A$_\mathrm{V}$ \citep{Glass1999book}. Such methods have been used for systematic studies of the extinction
in large areas from DENIS \citep{Schultheis1999} and 2MASS \citep{Dutra2003}, and also for modeling  of Galactic stellar populations from these surveys \citep{Marshall2006}.

When the luminosities of the red giants are known, e.g. those of the 'red clump' or the
RGB tip, one may infer both extinction and the
distance from colour-magnitude diagrams (CMD) such as $J$ vs $J-K_S$.
The numerous giants of the red clump are known
to be, by far, the best way to make 3D estimates of the extinction along Galactic lines of sight by
determining distance scales from their well defined luminosities and intrinsic colours.
\citet{Lopez-Corredoira2002}, \citet{Drimmel2003} and \citet{Indebetouw2005} have exploited the red clump stars of DENIS and 2MASS to explore the extinction along various lines of sight.

The 6-8~$\mu$m range is not as good an indicator of AGB mass-loss as 15~$\mu$m.
However, the sensitivity of ISOGAL at 7~$\mu$m allows the detection of less luminous giants
below the RGB tip at the distance of the bulge \citep{Glass1999}.
The large number of such stars,
$\sim$10$^5$, detected by ISOGAL in the Galactic disk and inner bulge may also be used to study the
mid-infrared extinction law at 7~$\mu$m by comparing $K_S$-[7] with $J-K_S$.  \citet{Jiang2003} have used this approach along one line of sight.
This method may even be extended to the derivation of extinction at 15~$\mu$m from the ratio $K_S$-[15]/$J-K_S$. It has recently been applied to all the exploitable ISOGAL lines of sight (more than 120 directions at both wavelengths of ISO) in the Galactic disk and inner bulge \citep{Jiang2006}.

Young stars with dusty disks or cocoons are the other class of objects to be addressed using ISOGAL's ability to detect 15 $\mu$m excess.
 \citet{Felli2002} have thus identified 715 candidate young stars from relatively bright ISOGAL sources with
a very red [7]-[15] color.  \citet{Schuller2002} has proposed another criteria for identifying such
luminous young stellar objects based on the non-point source like behaviour of the 15~$\mu$m emission. However,
various reasons have limited the full exploitation of ISOGAL for young star studies; e.g.  lack of
complementary data at longer or shorter wavelengths which makes it difficult to have a good diagnostic of the
nature of the objects, their luminosity and mass; lack of angular resolution which may preclude
deblending of nearby sources; limited quality of the data, especially in the regions of active
star formation with high diffuse infrared background.

Other infrared surveys, IRAS and MSX \citep{Price2001}, have covered the complete Galactic disk, including bands at longer wavelengths, but with a very limited sensitivity, especially in the range 12--20~$\mu$m where ISOGAL is three or four magnitudes deeper than MSX.
However, the much increased panoramic capabilities of {\it Spitzer Space Observatory} have now made
available much deeper data in the four IRAC bands (3.6, 4.5, 5.8 and 8.0$~\mu$m) in the main part of
the whole Galactic disk from the GLIMPSE {\it Spitzer Legacy Project} \citep{Benjamin2003,Benjamin2005}, extended to the whole inner disk with GLIMPSE II. GLIMPSE is about one
order of magnitude deeper than ISOGAL in the range 6-8$\mu$m, and it has a better angular
resolution, but it lacks extension at longer wavelength as provided by ISOGAL at 15~$\mu$m.
We note that the MIPSGAL project with {\it Spitzer} will provide longer wavelength (MIPS bands at 24$\mu$m and 70$\mu$m) coverage in the near future \citep{Carey2005}.

The main purpose of the present paper is to begin a reassessment of ISOGAL data in
conjunction with the availability of GLIMPSE data. We have chosen a standard ISOGAL field with good
quality data and covering a relatively large area and latitude range, allowing a valuable statistical study.
We have made a complete analysis of the ISOGAL data and validated their quality using the GLIMPSE data.
We discuss their various science outputs, especially at 15~$\mu$m, including complementary
information in the line of sight, especially from GLIMPSE/2MASS and \element{C}\element{O} millimetre observations. Our main goal from such a case study is to validate general methods for a subsequent complete exploitation of 15~$\mu$m data in all ISOGAL fields, complemented by near-IR and GLIMPSE data, especially for
systematic studies of AGB stars and their mass-loss and dusty young stars of intermediate mass.

		The paper is organized as follows: Sect. \ref{sec_dataset} recalls the general properties of ISOGAL
data and describes the associations of ISOGAL point sources in this field with GLIMPSE, 2MASS,
and MSX sources. Section \ref{sec_stats} discusses corresponding statistics and the validation of ISOGAL quality
using GLIMPSE data. Section \ref{sec_ext_dist} is devoted to interstellar extinction in this direction, the mid-IR extinction
law and stellar dereddening, and comparison with \element{C}\element{O} emission across the field, in order to infer a
rough three dimensional picture of extinction and source distribution. Section \ref{sec_nature} deals with the nature of the sources: the AGB population, its
mass-loss, luminosity function and relation with the RGB population, identification of the
relatively few young stars in this direction and their properties and relationship with
various other indicators of star formation.

\section{The data set}
\label{sec_dataset}

\subsection{ISOGAL observations and data reduction}

	The details of the ISOGAL observations with ISOCAM
\citep{Cesarsky1996} and the general processing of the data are described in \citet{Schuller2003}. The reduction of the data was performed first using the standard 'CIA' package
of ISOCAM pipeline \citep{Ott1999}, with post-processing specific to
ISOGAL data including  simulation of the time behaviour of the pixels of
the ISOCAM detectors; PSF source extraction optimised for crowded fields; removal of residual
source-saturation artefacts; specific photometric calibration; 7-15~$\mu$m band merging etc  \citep{Schuller2003}.
\begin{figure}
\includegraphics[width=\columnwidth]{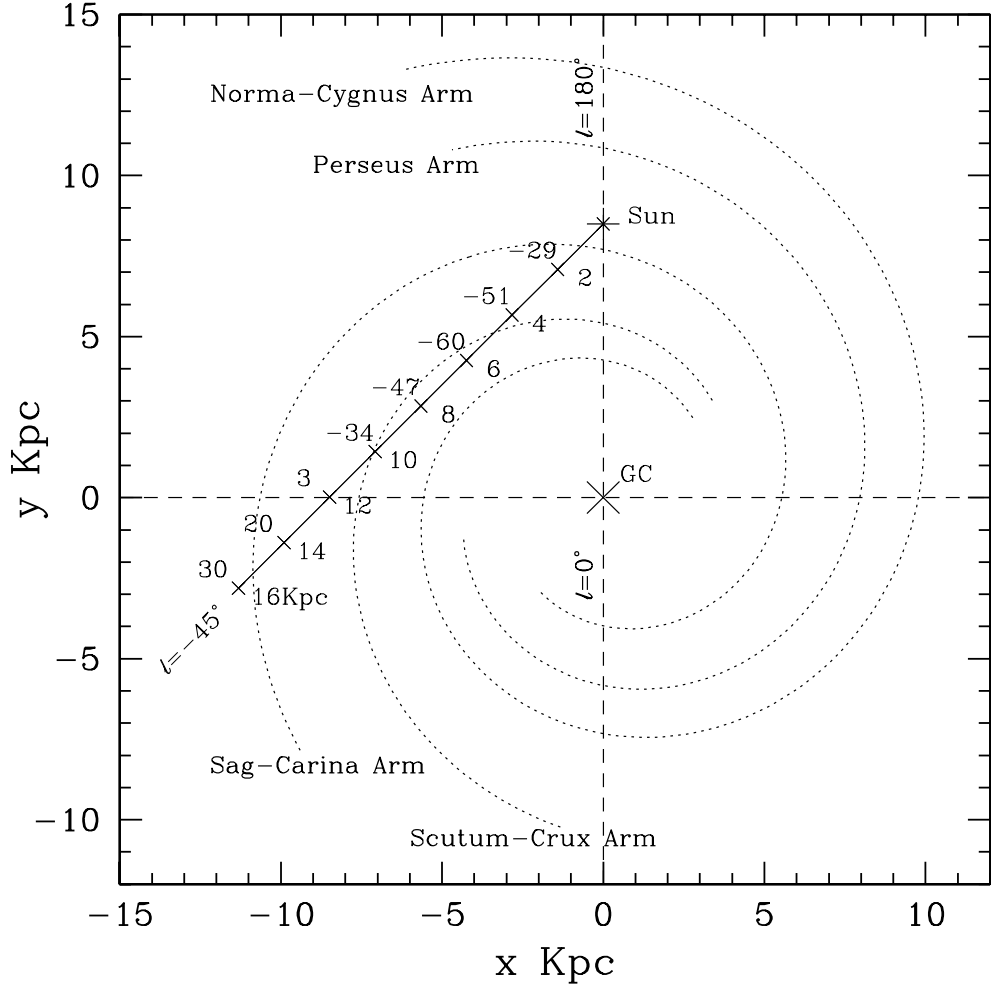}
\caption{The LN45 direction is shown as a solid line overlaid on the spiral arms (dotted curves)
from \citet{Russeil2003}.  Distances are marked every 2kpc below the solid line.  Numbers above the
solid line are the kinematic velocities in km~s$^{-1}$  of the \element{C}\element{O} clouds \citep{Lang_Book} at those distances.}
\label{spirals}
\end{figure}

\begin{figure}
\includegraphics[height=0.67\textheight]{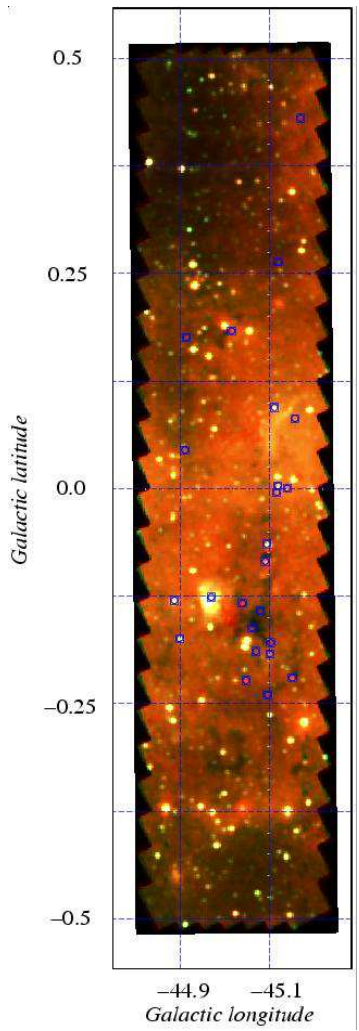}
\caption{False colour composite of the LN45 field with the 7$\mu$m images being set in the green and blue channels and the 15$\mu$m represented in the red channel to make up an RGB image. The individual channels have been scaled differently to bring out the extended emission: Red colour represents the 15$\mu$m image with histogram equalization while the green and blue colour represent the 7$\mu$m image linearly scaled between two different limits.    Galactic coordinate grid is overlaid.   Images were gaussian smoothed with a 2 pixel radius prior to composition.  Positions of the YSO candidates are marked with blue square symbols.}
\label{ln45image}
\end{figure}

For the present case study, we selected an ISOGAL field in the disk, with a large area (0.16 deg$^2$), LN45, towards longitude $\ell = -45\degr$,  covering the latitude range $\vert b \vert <  0.5\degr$.   The LN45 field is designated as FC-04496+0000 in the ISOGAL PSC \citep{Schuller2003} where FC represents fields observed in both 7$\mu$m and 15$\mu$m and the next digits represent the galactic longitude and latitude of the center of the field. The data are of good quality with  repeated observations (two observations at 15$\mu$m) for quality verifications. As shown in Fig. \ref{spirals}, this is in a direction tangential to the Scutum-Crux arm and also crosses the Sagittarius-Carina arm \citep{Russeil2003}.  A colour-composite image resulting from the ISOGAL observations is shown in Fig. \ref{ln45image}.  The young stellar object candidates (YSO, see Sect. \ref{sec_yso}) are marked in this figure with blue squares.
The observational parameters are reported in Table \ref{tab_obs_pars}.   A preliminary analysis of the same field, observed as part of the science verifications of ISOCAM, has been presented by \citet{Perault1996}.

Broad band ISOCAM filters at 7~$\mu$m ($LW2$) and 15~$\mu$m ($LW3$, Table 1) were used to increase
the sensitivity as in most ISOGAL observations.  Conversion factors from flux-density to
Vega magnitudes are:

	[7]  = 12.38 - 2.5 $\log F_{LW2}$ (mJy)

	[15] = 10.79 - 2.5 $\log F_{LW3}$ (mJy)

\begin{table*}[h]
\caption{Details of the ISOGAL observations discussed in the present paper}
\label{tab_obs_pars}
\begin{tabular}{ccclcccccr}
\hline
$\ell$&   $b$&     TDT& Observation date & ISO-ID& $\delta\ell/2$& $\delta$\,$b$/2& RA(2000)&
DEC(2000)& Filter\\
\hline
315.042&  +00.000& 24901257& 23 July 1996         &LN4500A& 0.105&  0.507& 218.067&  -60.477&  LW2\\
315.040&  +00.004& 30601587& 18 September 1996    &LN4500A& 0.125&  0.500& 218.044&  -60.481&  LW3\\
315.040&  +00.001& 60600458& 14 July 1997         &3N45P0 & 0.100&  0.502& 218.062&  -60.475&  LW3\\
\hline
\end{tabular}

Notes: i) The ISOCAM filter bandwidths (half-maximum transmission points), are 5.0-8.5$\mu$m for $LW2$ and 12.0-18.0$\mu$m for $LW3$ see \citep{Blommaert2003} \\
ii) $\delta\ell/2$ \& $\delta\,b/2$ are the dimensions in degree of the rectangular field retained for this study within the observed field, to avoid edge effects on the quality of data  and associations \citep{Schuller2003}\\
iii) Observations are with 6'' pixels.
\end{table*}

The standard ISOGAL-DENIS catalog for this field has been constructed using the ISO observations 24901257 and 60600458  in $LW2$ and $LW3$ respectively.  For the source identification and band-merging procedures used we refer to \citet{Schuller2003} and \citet{Omont2003}.
The number of sources extracted are 699 and 352 respectively in the $LW2$ and $LW3$
bands within the limiting magnitudes 9.87 and 8.62, corresponding to flux limits
of about 10 mJy and 7 mJy, respectively. In the ISOGAL-DENIS point source catalog (PSC) there are thus a total of 746 ISOCAM sources, out of which 305 objects are detected in both $LW2$ and $LW3$ bands, 394 objects detected in only $LW2$ and 47
objects detected only in $LW3$ band.  Among the  305  $LW2$ -- $LW3$ associated sources, 289
sources  have good association quality flags (3 or 4)  and  16 sources have doubtful
associations with quality flag 2 \citep{Schuller2003}.   With an association radius of 5.4\arcsec and a one sigma distance uncertainty of  1.7\arcsec, the upper limit to the number of chance associations between $LW2$ -- $LW3$ is less than 1 (0.3\% of 305 common sources).

The dedicated DENIS data for this field,  as published in the version 1 of the ISOGAL--DENIS PSC catalog, are extracted from a series of observations in 1996 and 2000.   We have complemented this set with additional DENIS data from the regular DENIS observations (strips) for the whole field.  The identification procedures employed are the same as discussed in detail by  \citet{Schuller2003}.

\subsection{Cross identification with other surveys}

Sources in the ISOGAL--DENIS catalog were cross-identified with corresponding sources in the other
large scale surveys -- 2MASS, GLIMPSE, and MSX.   The procedure adopted is discussed below.
At all stages the process was tailored to reduce the possibility of chance associations.

\subsubsection{Association between ISOGAL--DENIS and GLIMPSE--2MASS}
\label{assoc_ID_G2}
Version 2 of the GLIMPSE catalog{\footnote{see \url{http://data.spitzer.caltech.edu/popular/glimpse/20070416_enhanced_v2/source_lists/}}} provides the 2MASS photometry for the GLIMPSE counterparts.  The total number of sources extracted in the $IRAC4$ band of the GLIMPSE--2MASS catalog is 4601.  We note that sources saturated in GLIMPSE but not saturated in 2MASS are not included in the version 2 of the GLIMPSE--2MASS catalog.  The saturation limit in GLIMPSE $IRAC4$ band is 4 mag.

The Spitzer pointing accuracy is better than 1\arcsec \citep{Werner2004, McCallon2007}.  The ISOGAL/DENIS catalog PSC has its astrometry derived from that of the DENIS $I$ band which is in turn based on the USNO catalog. Therefore the positional uncertainty in the case of the ISOGAL PSC is also of the order of 1\arcsec.

The sources in the ISOGAL--DENIS catalog were cross-identified with the GLIMPSE--2MASS catalog using
search radii of 3.8\arcsec.  To have probability of chance association below 10\% the search radius, $r_{s}$, was computed from the requirement that $n \pi r_{s}^2 = 0.1$, where $n$ is the source density of the GLIMPSE--2MASS catalog.  Here $n=28756$ sources per sq. deg giving $r_{s}=3.8$\arcsec.   The probability of false associations drops to less than 4\% when one considers that 95\% of the ISOGAL sources are associated to a GLIMPSE--2MASS source within 2.3\arcsec.
Based on the association distance and various photometric criteria we define a set of quality flags as listed in table \ref{flagstats}.

\begin{figure}
\includegraphics[width=\columnwidth]{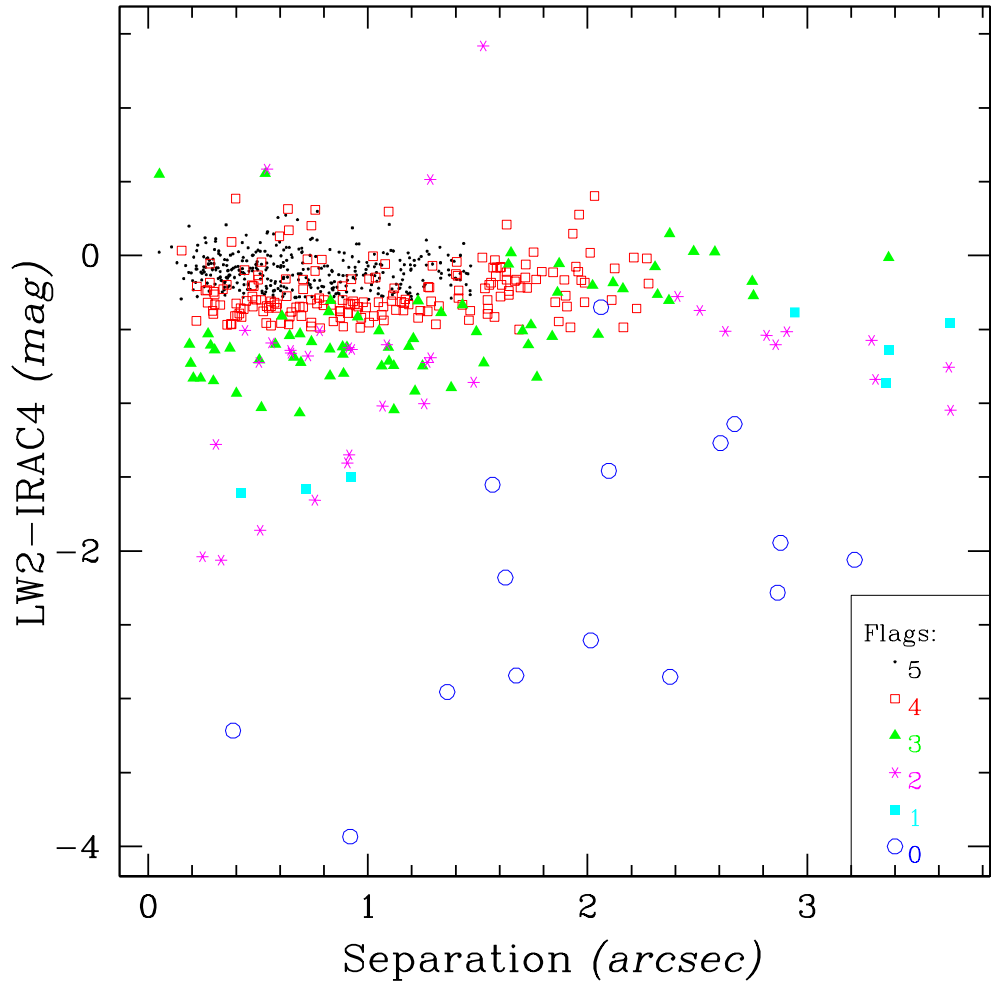}
\caption{The relative reliability of the sources as a function of distance of association and 7$\mu$m-8$\mu$m colour.  This is used to define the various flags (from 0 to 5).}
\label{flagdefs}
\end{figure}

The definition of the association or quality $flags$ (based on a combination of photometry and astrometry) is described in detail in the {\em readme} file accompanying the electronic version of the LN45 catalog and also in the online appendix (see Sect. \ref{flag_readme}) with this article.  Fig. \ref{flagdefs} illustrates the $flag$ definitions for the cases where associations exist between GLIMPSE and ISOGAL (on a scale of 1 to 5).
We list the total number of sources with the various $flags$ in our catalog in Table \ref{flagstats}.
Sources with a $flag$ of 0 (cases of weak detections in only one of the ISO bands) are excluded from the discussion.  The final catalog for the LN45 field contains 681 sources (after excluding the $flag$ 0 sources) out of which 8 sources do not have
GLIMPSE counterparts in the $IRAC4$ band within 3.8\arcsec.   Three of these are cases due to saturation in the 8$\mu$m band ($flag$ = 9).  Five relatively bright  sources found in the ISOGAL catalog but not in the GLIMPSE 8$\mu$m (but below the saturation limits of GLIMPSE) are apparently extended sources or blends.  There are no 8$\mu$m sources within 5\arcsec of the ISOGAL source in the GLIMPSE survey for three of these sources ($flag$ = 8), while the other two have GLIMPSE counterparts at 4.2\arcsec ($flag$ = 7).

\subsubsection{Associations with MSX}

The MSX catalog in the LN45 field area contains 76 sources,  of which only 13 are detected in band $E$ (21.34$\mu$m); 36 are detected in band $D$ (14.65$\mu$m); 37 in band $C$ (12.13$\mu$m) and 76 in band $A$ (8.28 $\mu$m).   With positional uncertainties of the order of 1\arcsec \citep{Egan2003} 70 of the MSX sources (92\%) are associated with ISOGAL sources within 5\arcsec.   Two sources having dubious MSX quality flags are associated with rather faint ISOGAL sources.

\subsubsection{Multi-band Point Source Catalog}

\begin{table*}
\caption{The different association flags and number statistics}
\label{flagstats}
\begin{tabular}{crl}
\hline
Flag & Number of sources & Remarks\\
\hline
9&3&ISOGAL sources saturated in GLIMPSE $IRAC4$\\
8&3&ISOGAL sources of intermediate brightness missing in GLIMPSE\\
7&2&ISOGAL sources of intermediate brightness with GLIMPSE counterpart at $r > 3.8$ and $r < 4.5$\\
5&355&Very secure association between ISOGAL and GLIMPSE\\
4&188&Secure association between ISOGAL and GLIMPSE\\
3&80&Probable association\\
2&41&Possible association\\
1&9&Dubious association, probably real but blended sources\\
0&65&Rejected sources\\
\hline
total&746&Number of sources in the ISOGAL-DENIS PSC version 1 \citep{Schuller2003}\\
\hline
\end{tabular}
\end{table*}

Finally we have a catalog of 673 reliable ($flag=1$ to $flag=5$) ISOGAL sources with measurements in the GLIMPSE survey.    We also include the 8 additional sources discussed above (with $flags$ 7--9) to get a total of 681 sources.  MSX, 2MASS and DENIS photometry is provided where available.    A brief extract of the table is in the Online Sect. \ref{online_table}. The entire table is available electronically at the CDS.  The format of the catalog with the details of the columns is listed below.

\begin{itemize}
 \item identification number
 \item right ascension
 \item declination (from GLIMPSE-2MASS) else ISOGAL-DENIS if no GLIMPSE-2MASS
 \item ISOGAL-DENIS PSC1 name
\item DENIS $I$, $J$, $K_S$
 \item 2MASS $J$, $H$, $K_S$
 \item $IRAC1$ to $IRAC4$
 \item $[7]$
 \item $[15]$
 \item MSX band $A$ to $E$
\item ISOGAL-GLIMPSE association flag
\item type of source (AGB, RGB, YSO, PNe)
\item distance (kpc) and $A_\mathrm{V}$ (mag) as derived assuming the Red Clump Locus method (see Sect. \ref{distdeterm}) with typical errors.
\item $\mathrm{M}K_S$ and $\dot{M}$ for the AGB/RGB (see Sect. \ref{sec_massloss})
\item general remarks about the source including comments on validity or otherwise for derived distance.
\end{itemize}

\begin{figure*}
\includegraphics[width=\textwidth]{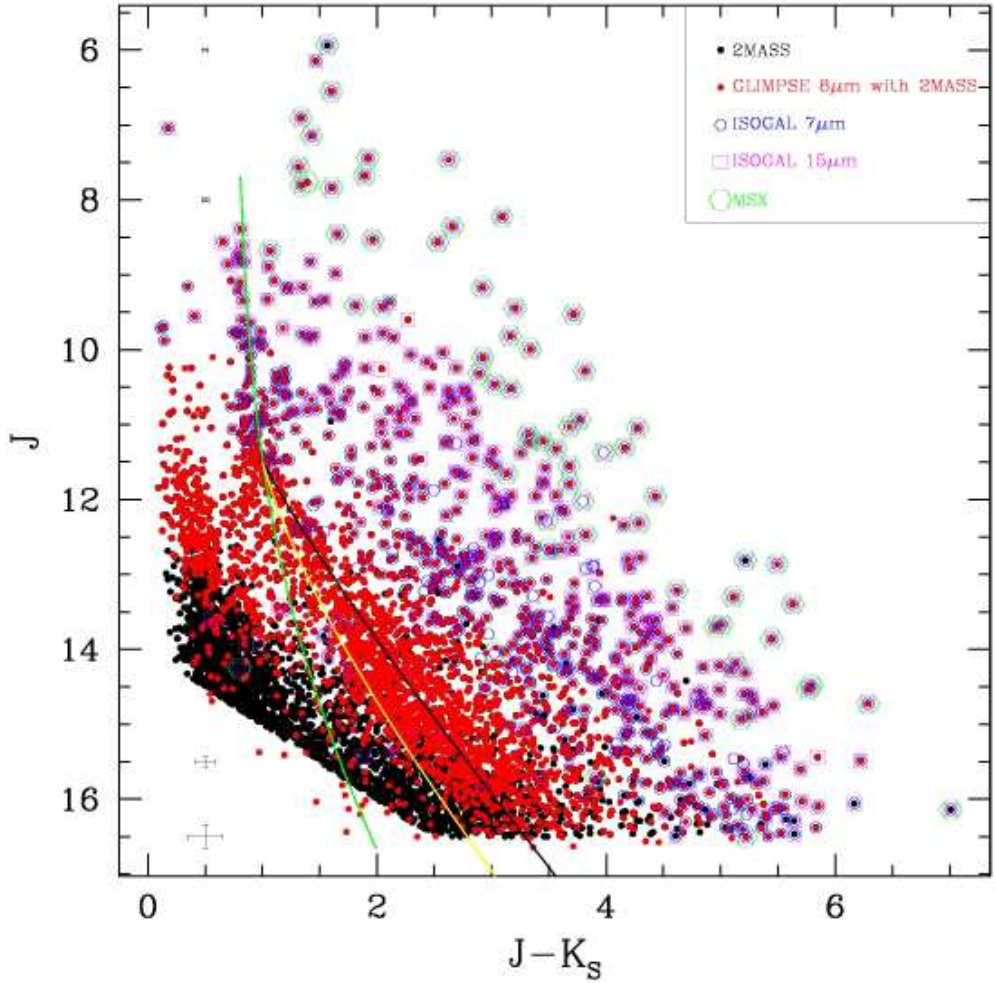}
\caption{The $J-K_S$ vs $J$ CMD.
Overlaid are symbols for detections in other surveys as shown in the legend (top
right corner).  The red-clump locus (RCL) with normal extinction is shown as a
solid green curve.  The yellow and black curves mark RCL with larger extinction
per unit kpc (see Sect. \ref{sec_ext_dist}).  Typical error bars are marked at
$J$=16.5,15.5,8 and 6.0 at $J-K_S$=0.5.  Online material includes separate figures with sources selected based on detections in the various surveys.}
\label{jjk_surveys}
\end{figure*}
In Fig. \ref{jjk_surveys}, we show the 2MASS CMD $J-K_S$ vs $J$ with the associated mid infrared sources from different surveys (see also the online Sect. \ref{jjk_cmd_online}).   Note the varying levels of completeness in the various surveys.  The completeness is a function of the wavelength, sensitivity and the spatial resolution of the surveys.   This diagram will be discussed in detail in Sect. \ref{sec_ext_dist} for source distance and extinction determination.

\section{Accuracy of ISOGAL data} 
\label{sec_stats}

In this section, we discuss the accuracy of the ISOGAL data using repeated observations (at $15\mu$m)
and using the completeness of identification with GLIMPSE.

\subsection{Comparison with GLIMPSE data}

\begin{figure*}[ht]
\includegraphics[width=0.52\textwidth]{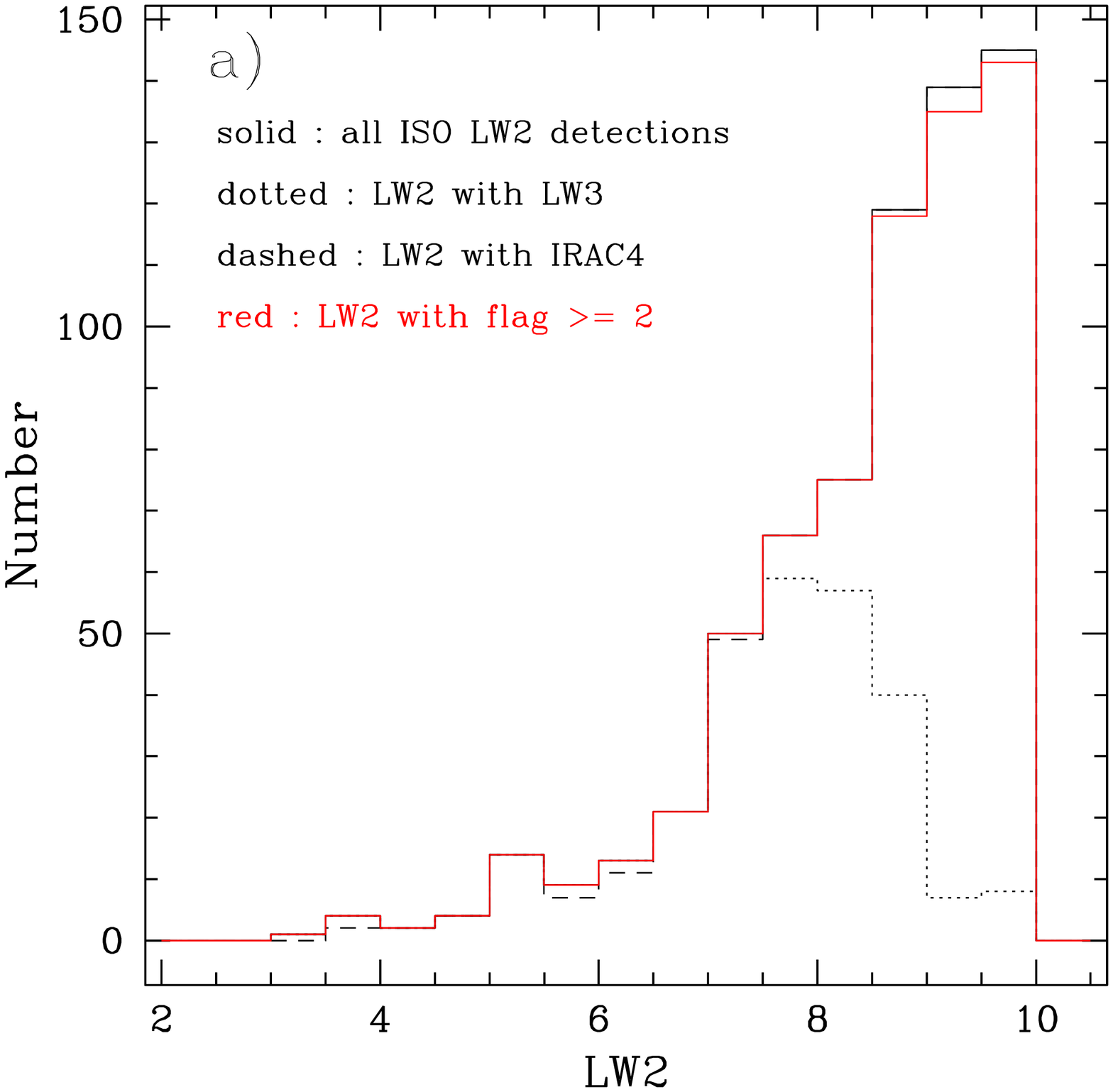}
\includegraphics[width=0.52\textwidth]{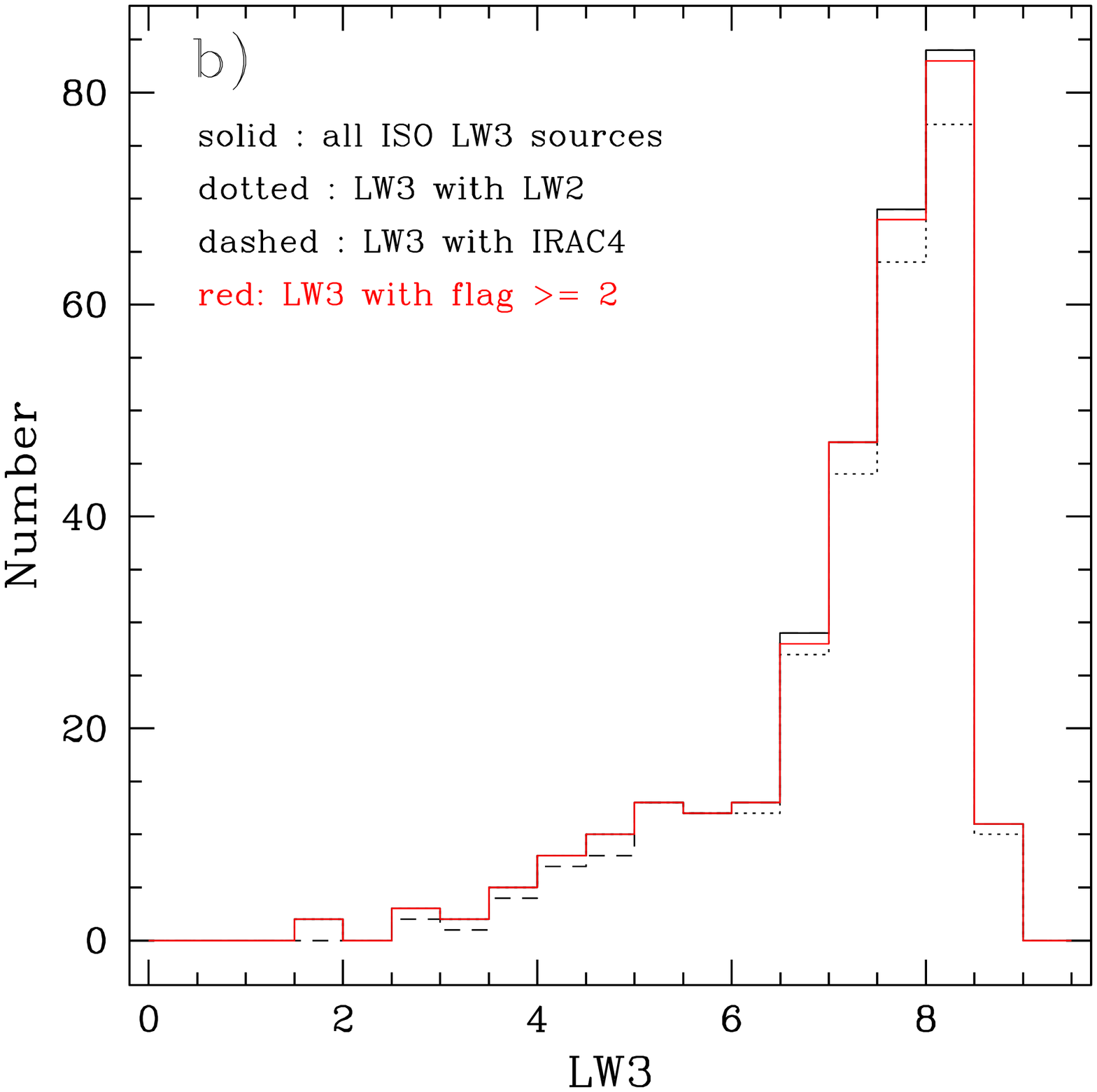}
\includegraphics[width=0.52\textwidth]{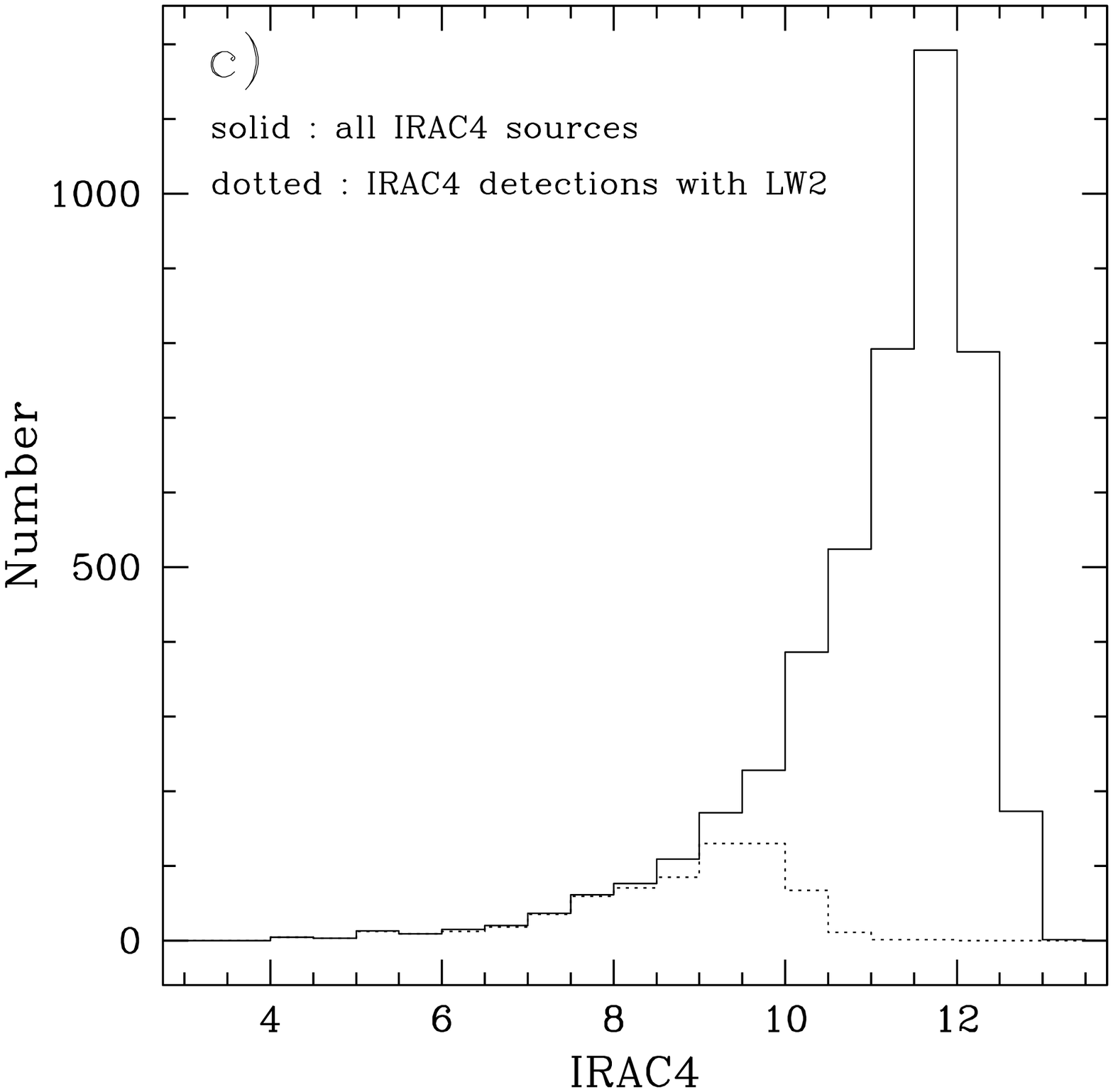}
\caption{
(a) The histogram of the 7$\mu$m sources binned by magnitude (binsize = 0.5mag) detected in this field with detections also at 15$\mu$m (dotted line); detections in GLIMPSE 8.0$\mu$m (dashed).  In red are shown the 7$\mu$m sources with a $flag \ge 2$.  (b)  The histogram of the 15$\mu$m sources detected in this field with detections also in 7$\mu$m (dotted), and in 8.0$\mu$m in GLIMPSE (dashed).  15$\mu$m detections with $flag \ge 2$ are shown in red.  (c) The histogram of the GLIMPSE 8.0$\mu$m sources detected in this field with detections also in LW2 (7$\mu$m) (dotted).}
\label{histo}
\end{figure*}

\begin{figure*}
\includegraphics[width=0.75\textwidth]{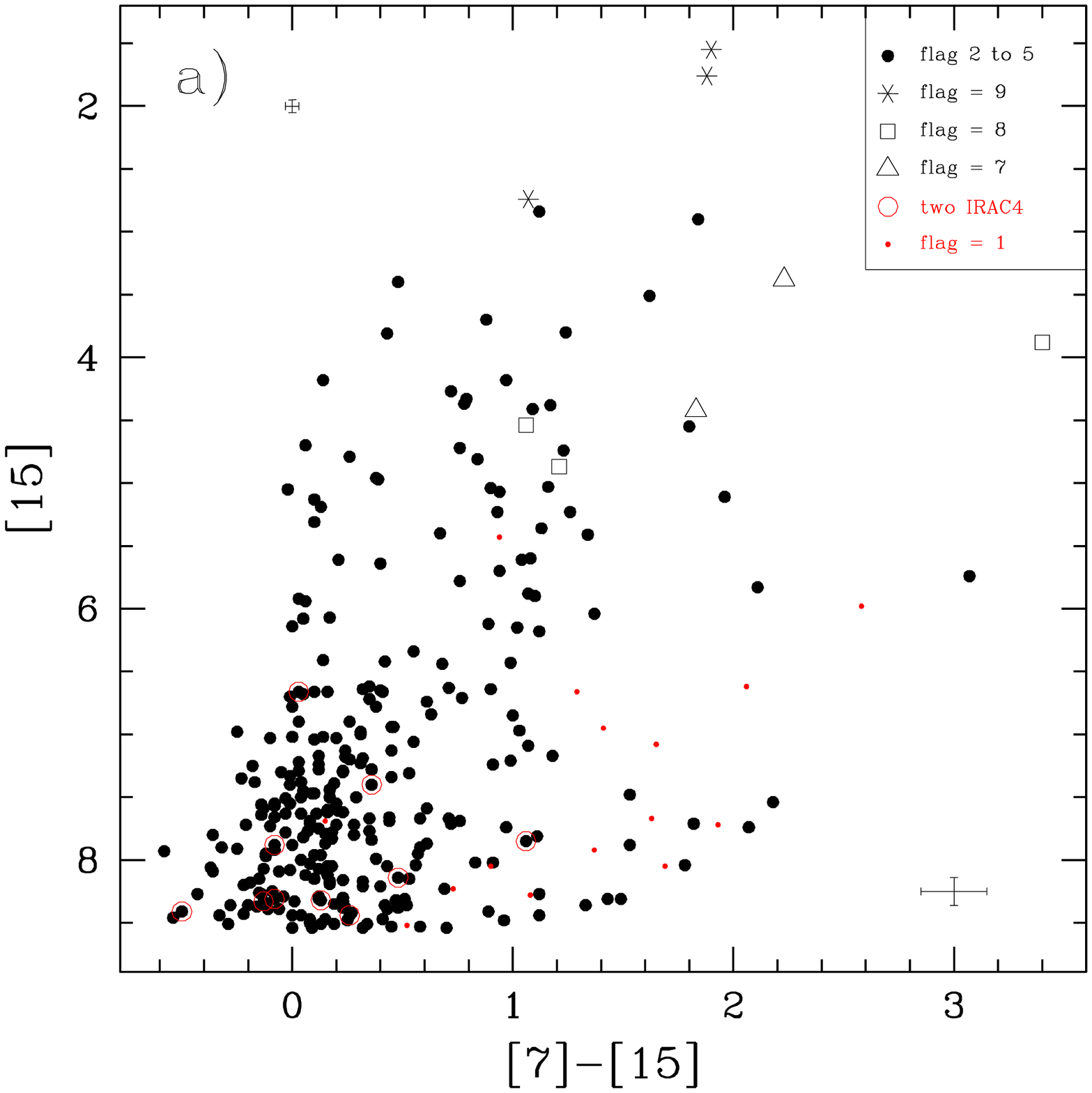}
\includegraphics[width=0.75\textwidth]{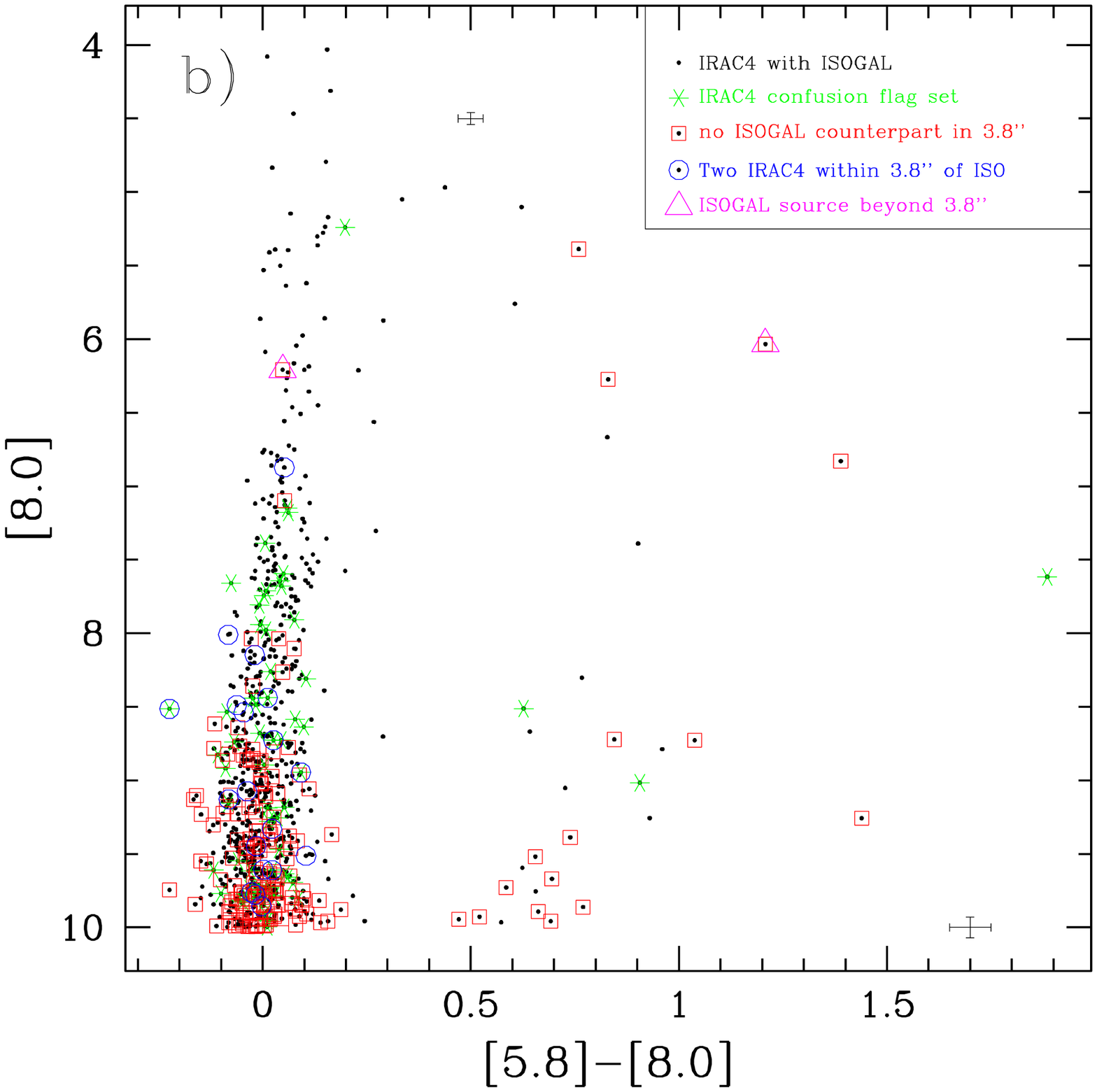}
\caption{
(a) ISOGAL CMD. The sources missed by GLIMPSE due to saturation are marked by asterisk.  Open triangles mark the ISOGAL sources matched to GLIMPSE sources at larger distance (4.2\arcsec) i.e. $flag=7$.  Blended/extended ISOGAL sources not in GLIMPSE are shown as open squares.  Sources with a second GLIMPSE source within 3.8\arcsec are shown with large open circles.
(b) GLIMPSE CMD showing the sources that are missed by ISOGAL (open squares).  We have limited the display to bright GLIMPSE sources ($IRAC4$ brighter than 10).}
\label {gl_isofig}
\end{figure*}

In Fig. \ref{histo} we exhibit the number counts versus magnitude histograms of (a) the 7$\mu$m detections (b) the 15$\mu$m detections, and (c) the IRAC4 8$\mu$m detections.  They show that the ISOGAL detections are 94\% complete up to [7]=8.5$^m$ (GLIMPSE 8.0$\mu$m magnitude) for the LW2 detections when compared to the GLIMPSE 8$\mu$m.  The 15$\mu$m data are complete upto [15]=7$^m$ when compared to ISOGAL 7$\mu$m.

We find that, apart from the 65 rejected sources (8.7\% of 746, see section \ref{assoc_ID_G2}), most  of the ISOGAL sources (98.8\% of 681) are detected in GLIMPSE.

Three ISOGAL sources below the GLIMPSE saturation limits do not have 8$\mu$m  counterparts ($flag$ of 8 and 7) within 3.8\arcsec.   However, going to larger association radius retrieves GLIMPSE counterparts for two of the ISOGAL sources (with a corresponding flag of 7).  Inspection of the images show that these 5 sources are slightly extended (blended) in the ISOGAL images while they are resolved by GLIMPSE into individual sources.   These 8 sources are shown with different symbols in the ISOGAL CMD of Fig. \ref{gl_isofig}a.

A GLIMPSE colour magnitude diagram is shown in Fig. \ref{gl_isofig}b where we overplot the ISOGAL
non-detections with different symbols.  As can be seen, only a small fraction of the bright sources are missed by ISOGAL due to poorer resolution (blends).  These sources are shown as open squares in Fig. \ref{gl_isofig}b.

\citet{Schuller2002} had found that extended sources in the ISOGAL inner bulge fields had relatively much larger photometric errors (derived from the fitting of the point spread function).
We note that the blended/extended sources in the LN45 field do not show any similar trends.

\subsection{Repeated 15$\mu$m ISOGAL observations}

Repeated ISOGAL 15$\mu$m observations show that the dispersion (rms) in the 15$\mu$m photometry is better than 0.3$^m$ for the bulk of the sources while it reaches 0.5 at the fainter end.  Three sources are found to be variable in the two epochs of the ISOGAL observations and are discussed in the section \ref{sec_massloss}.

\section{Interstellar extinction and distance scale} 
\label{sec_ext_dist}

\subsection{Local value of optical extinction from main-sequence stars}

Field LN45 is located across the Galactic mid-plane where the interstellar
extinction is expected to be high. The measurement of interstellar  extinction in the optical range
is possible only for the nearby regions.  \citet{Neckel1980} have reported interstellar extinction $A_\mathrm{V}$ based on the optical observations of O-type to F-type stars in the direction of $\ell=314\degr$ and
$b=0.0\degr$, which is close to LN45 field, to be nearly uniform between 2 and 3
magnitude up to a distance of $\sim$ 3.5kpc. This ISOGAL field direction, as already mentioned, crosses the
Sagittarius-Carina arm at 1-2~kpc and 15~kpc, and it is almost tangential to the Scutum-Crux arm
from $\sim$4 to 10~kpc (Fig. \ref{spirals}). The stars considered in the study  by \citet{Neckel1980}
belong either to the Sagittarius-Carina arm or the nearby region in front of this arm or just
beyond. The determination of the extinction to farther distances must rely on the
near-infrared, which is much less absorbed, and the reddening of red giants.

\subsection{Near-infrared CMDs of red giants, extinction and distance scales}
\label{distdeterm}

\begin{figure*}
\includegraphics[width=0.67\textwidth]{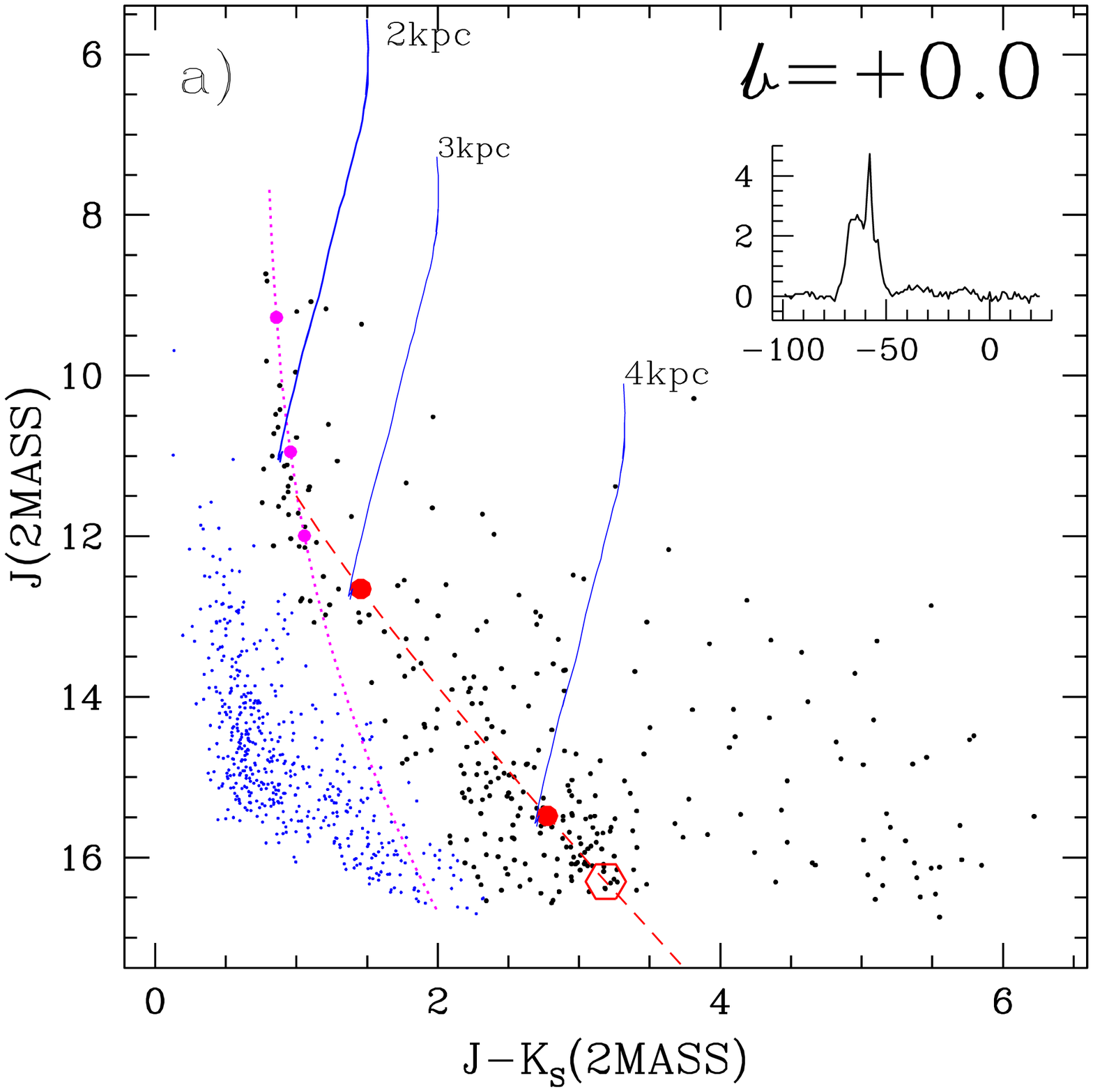}
\includegraphics[width=0.67\textwidth]{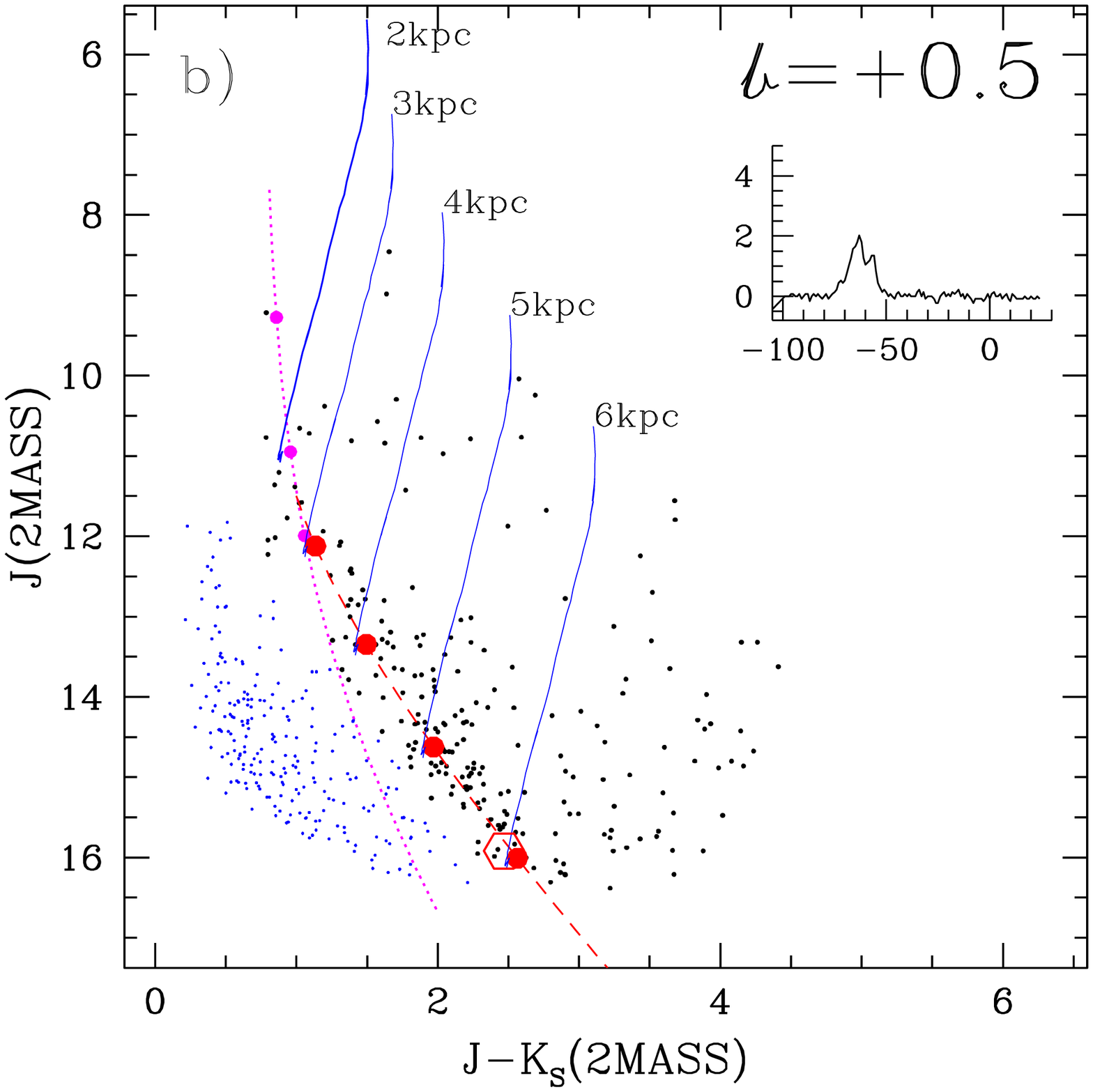}
\caption{
(a) $J$ vs $J$-$K_S$ CMD.   Blue dots are mostly foreground sources.   Black dots are red clump giants and other mid-infrared stellar populations.  The red clump locus with normal or uniform extinction  (cJ=0.166) marked as dotted (magenta) line; extinction as a function of distance marked as a dashed (red) line.  This is for the range $-0.0625 < b < 0.0625$. i.e. $b=0.0$ of Table \ref{tab_av_dist_cmd_co}.  Also shown are the RGB/AGB isochrones at distances of 2kpc (with normal extinction law), and at 3 and 4kpc (modified law) as labeled in the figure.  We use  $A_\mathrm{V}=6.0(A_\mathrm{J}-A_\mathrm{K_S})$ (\citet{Glass1999book}, and Table \ref{tab_akl_ajk}).
(b) CMD for the range $0.4375 < b < 0.5$ i.e. $b=0.5$ of Table \ref{tab_av_dist_cmd_co}.  All other details are as mentioned in (a). Additional isochrones at 5 and 6kpc are also shown.  In both panels, we show with large open hexagons the points listed in table \ref{tab_av_dist_cmd_co}.  The corresponding \element{C}\element{O} spectra are also shown as an inset with the velocity scale (in km/sec) marked on the x axis.  See the electronic (online) version of the paper for plots corresponding to other values of $b$.
}
\label {jjk_dist}
\end{figure*}

Fig. \ref{jjk_dist} shows the most useful diagram for the purpose of the determination of the distance and extinction in the direction of LN45 - the CMDs $J-K_S$ vs $J$ towards two
different ranges in galactic latitude as mentioned in the captions.  The
red clump sequence is clearly seen in Fig. \ref{jjk_dist}a and \ref{jjk_dist}b.  The red clump sequence can be identified at $J$ of $11^m$ and $J-K_S$ of $~1$. Going towards fainter $J$ their $J-K_S$ also becomes redder and at the base of the CMD reaches $J-K_S$ beyond $2.5^m$.  As discussed e.g. by \citet{Indebetouw2005} for a similar line of
sight \citep[see also][]{Lopez-Corredoira2002}, the red clump giants can be isolated and their
distribution fitted to determine the extinction with distance. In the diagrams of Fig. \ref{jjk_dist},
 the red clump giants are well isolated from the foreground main sequence stars,
mostly numerous, faint, cool dwarfs, and the background more luminous giants up to the RGB tip and
the upper AGB. The latter are discussed below in Section \ref{sec_nature}.  Note that the analysis in this section is based on the entire 2MASS catalog for this field.

In a large field such as LN45, the extinction may vary significantly because of the range of galactic
latitudes and the clumpiness of molecular clouds. With the very large number of clump stars
available, one can try to use them to trace the extinction in a number of smaller, more homogeneous
spatial cells tiling the whole field.

In order to define an approximate distance scale, we first assume that the average extinction
in each field is just proportional to the distance $d$, and fit a mean curve for the red clump locus
(RCL) following the method of \citet{Indebetouw2005} and their Eqs. 2 \& 3:

 \begin{equation}
 	J = M_{J} + 5\log(d/10pc) + c_\mathrm{J}(d/10pc)
 \end{equation}

 \begin{equation}
 	J - K_S = M_J - M_{Ks} +(c_\mathrm{J}-c_\mathrm{Ks})(d/10pc)
 \end{equation}

where $c_\mathrm{J}$ and $c_\mathrm{Ks}$ are the average extinction per unit distance in the $J$ and $K_S$ bands. We assume
the same standard values as used by \citet{Indebetouw2005} and \citet{Lopez-Corredoira2002} \citep[see also][]{Wainscoat1992} for the mean absolute magnitudes of clump stars : $M_\mathrm{J}$ =
-0.95, $M_\mathrm{Ks}$ = -1.65, $J_\mathrm{0}$-K$_{S\mathrm{0}}$ = 0.75, and  extinction ratios : $c_\mathrm{J}$/$c_\mathrm{Ks}$ = $A_\mathrm{J}$/$A_\mathrm{Ks}$ = 2.5. The curve of the red clump locus then depends only on the extinction parameter $c_\mathrm{J}$ to be fitted.

However, it is clear from Fig. \ref{jjk_dist} that such a simple curve (dashed) with a single value of $c_\mathrm{J}$ for each subfield cannot properly fit the actual red clump locus. This is not surprising since the
variation of extinction with distance, represented by $c_\mathrm{J}$, is much smaller (by factors $\sim$2 to
7) locally than within the Scutum-Crux spiral arm, especially where the molecular clouds are located.

For defining such cells, it seems logical to use the galactic latitude, $b$, limits corresponding to the beams of the \element{C}\element{O} survey of \citet{Dame1987}, in order to make a correlation with the specific extinction of the molecular clouds inferred from the \element{C}\element{O} intensity.
Nine \element{C}\element{O} spectra cover nearly completely the LN45 $b$ range and we have defined nine cells corresponding to the $b$ range of each beam.  These are  tabulated in Table \ref{tab_av_dist_cmd_co}.
The \element{C}\element{O} spectra are inset in Fig. \ref{jjk_dist}.

For each cell, we have got a reasonable fit of the global RCL with the following procedure. We first determined the best value of $c_\mathrm{J}$ for the two ends of the curve: the local one, and the one for the largest extinctions available in each group corresponding to some location in the Scutum-Crux arm.
The combination of these two segments provide a reasonable fit for more than half of the curve in
all cases.  By inspection we determine the value of $J-K_S$ (or of $d$ or $A_\mathrm{V}$) for which the considered
segment is an acceptable approximation; and we finish by joining the two segment ends by a linear
interpolation in $c_\mathrm{J}$.  The distance and $A_\mathrm{V}$ at the farther end of the RCL and the corresponding $J$, $J-K_S$ are listed in Table \ref{tab_av_dist_cmd_co} along with the $A_\mathrm{V}$ derived from the \element{C}\element{O} strength.   Values obtained from the Besan\c{c}on model of the Galaxy \citep{Robin2003} are also included in this table for comparison. All the values of $c_\mathrm{J}$ for the local segments are the same ($c_\mathrm{J}=0.166$).   The uncertainties in $d$, $c_\mathrm{J}$ and $c_\mathrm{Ks}$ are of the order of 500pc, 0.1mag/kpc and 0.03mag/kpc respectively.

\begin{figure*}
\includegraphics[width=0.67\textwidth]{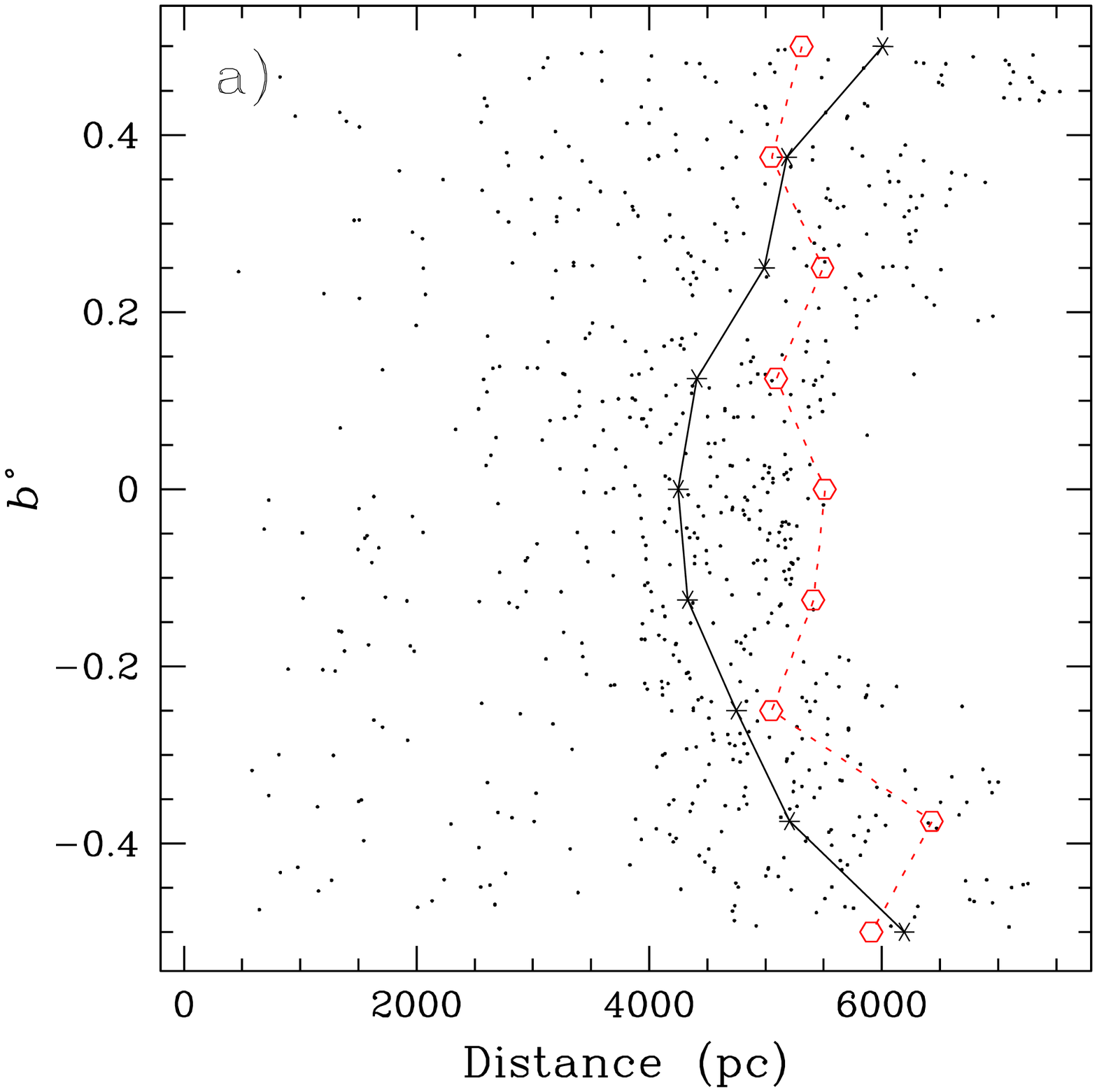}
\includegraphics[width=0.67\textwidth]{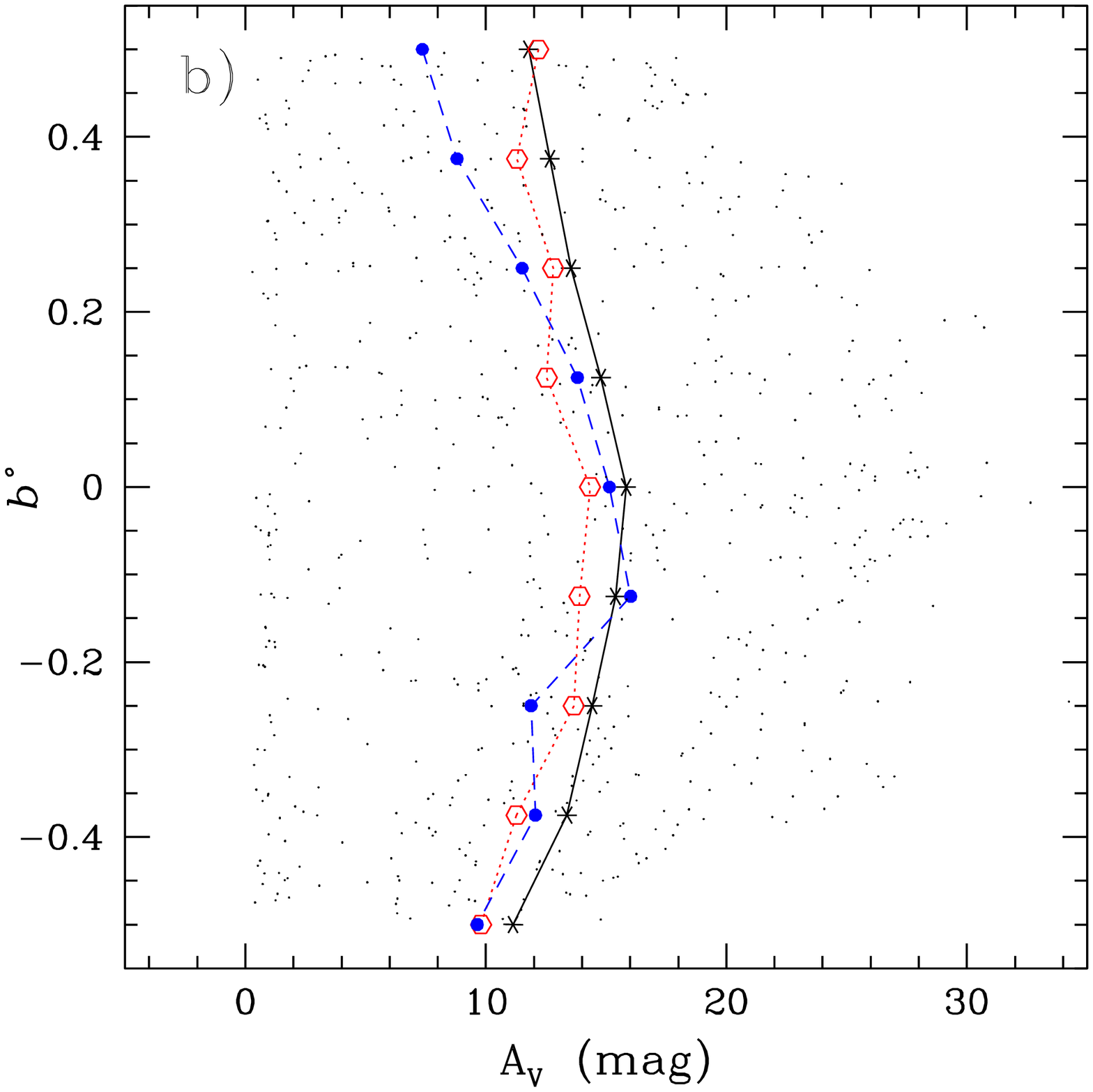}
\caption{ (a) Plot showing the derived positions for the sources as a function of galactic latitude $b$.  The solid line (black) connecting the asterisks represents the location of the red clump at the base of the CMD.   The dotted line (red) represents the corresponding red clump location from the Besan\c{c}on model \citep{Robin2003}. (b) Derived $A_\mathrm{V}$ as a function of galactic latitude.  The solid line (black) represents the Av at the position of the red clump in the observed CMDs.  The dotted line (red) represents the values from the Besan\c{c}on model while the dashed line (blue) is the $A_\mathrm{V}$ derived from the $^{12}$\element{C}\element{O} as listed in Table \ref{tab_av_dist_cmd_co}. }
\label{dist_lat}
\end{figure*}

Table \ref{tab_av_dist_cmd_co} and Fig.  \ref{dist_lat} compare the extinction and
distances derived with those from the Besan\c{c}on model.
The Besan\c{c}on model implements the \citet{Marshall2006} 3-D extinction map.
We note that the spatial resolution of the \citet{Marshall2006} map is $15\arcmin \times 15$\arcmin.
Fig. \ref{dist_lat}  shows small discrepancies between our derived extinction values and distances
compared to the values from the model.   However, considering the fact that the model relies on
many parameters such as scale length and  height of the thin and
thick discs, metallicity distribution, etc, the discrepancies are not significant.

For the highest latitudes, $\vert b \vert$~$\sim$~0.5$\degr$, it is seen that the
distances traced by the RCL range up to 7~kpc (see Fig. \ref{dist_lat}a), i.e. over a substantial part of the Scutum-Crux arm.
However, at the lowest latitudes, $\vert b \vert$~$<$~0.2$\degr$, only red clump stars in the nearest part of the Scutum-Crux arm, at 4-5~kpc, are detected in
$J$ by 2MASS.  The largest $A_\mathrm{V}$ seen in this field is close to 30 magnitudes for stars close to the RGB tip with $J-K_S$ $\sim$ 6 magnitude.

\begin{table*}
\caption{$A_\mathrm{V}$ (mag) and distance (pc) derived from red clump at the bottom end of the $J$ vs $J$-$K_S$ CMD.  Also shown are corresponding values of $J$, $J-K_S$, and the $A_\mathrm{V}$ derived from the \element{C}\element{O} spectral survey of \citet{Dame1987} and the corresponding $A_\mathrm{V}$ and  distance from the Besan\c{c}on Model (BM) \citep{Robin2003}.  All the cells are taken centered at $\ell=315.0$ between the two $b$ values ($b_{low},b_{high}$) shown.}
\label{tab_av_dist_cmd_co}
\begin{tabular}{cccccccccccc}
\hline
$b$&$b_{low}$&$b_{high}$&$J$&$J-K_S$&$A_\mathrm{V}$&Distance&$A_\mathrm{V}$(\element{C}\element{O})&$J$ (BM)&$J-K_S$(BM)&$A_\mathrm{V}$(BM)&distance(BM)\\
\hline
0.5&     0.4375&   0.5&      16.02&  2.57&  11.78&  6008&   7.36&  16.08&  3.03&  12.18&   5310\\
0.375&   0.3125&   0.4375&   15.92&  2.70&  12.66&  5184&   8.80&  15.93&  2.86&  11.30&   5050\\
0.25&    0.1875&   0.3125&   16.07&  2.84&  13.54&  4990&  11.51&  16.04&  3.07&  12.80&   5490\\
0.125&   0.0625&   0.1875&   16.11&  3.03&  14.78&  4410&  13.81&  16.05&  3.00&  12.53&   5090\\
0.0&     -0.0625&  0.0625&   16.30&  3.19&  15.83&  4250&  15.14&  16.21&  3.40&  14.34&   5510\\
-0.125&  -0.1875&  -0.0625&  16.23&  3.12&  15.39&  4331&  16.02&  16.22&  3.30&  13.90&   5410\\
-0.25&   -0.3125&  -0.1875&  16.19&  2.98&  14.42&  4750&  11.88&  16.25&  3.22&  13.66&   5050\\
-0.375&  -0.4375&  -0.3125&  16.11&  2.81&  13.37&  5207&  12.06&  16.08&  2.76&  11.27&   6430\\
-0.5&    -0.5&     -0.4375&  15.92&  2.47&  11.13&  6194&   9.64&  15.92&  2.55&   9.80&   5910\\
\hline
\end{tabular}
\end{table*}

In the various CMDs, e.g. $J$ vs $J-K_S$, for each distance (and reddening) determined from the
red clump, one may trace the corresponding isochrone of the red giants of various
luminosity (see e.g. Fig. \ref{jjk_dist}), assuming it is uniquely defined and in particular that these stars
have approximately the same age and metallicity in a given direction. For each ISOGAL red giant,
including most AGB stars without too strong mass-loss, one may thus determine its distance and
luminosity by tracing the corresponding isochrone \citep{Bertelli1994} from the position of the source in the plot and considering its intersection with the red clump locus (Fig. \ref{jjk_dist}).
This method applies to the majority of ISOGAL sources since they all are such red giants.
In the absence of other information about the nature of the source, we have used it systematically for all ISOGAL sources, at least as a first approximate step. However, it should obviously be modified and adapted to the actual nature of the sources, in particular for the three other main classes apart from the red giants:

1) Main sequence stars; ISOGAL detects them only at very
short distances (e.g. $\sim$ 100 pc for a K0V star, $\sim$ 250 pc for an F3V star, $\sim$ 300 pc for
a B9 star). They have thus small reddening, and bluer intrinsic colours or smaller luminosity than
red clump stars. The most obvious cases among ISOGAL sources are a few sources located below the
red clump locus or to the left of the $A_\mathrm{V}$=0 isochrone of red giants in Fig. \ref{jjk_dist}.

2) Young stars; class II and III protostars have
intrinsic near-IR colours not very different from that of red giants, so that
de-reddening them as if they were red giants, may give sufficient constraints on
their distance and luminosity.  However, being embedded in dust adds to the
uncertainty in estimating the extinction and distance to many of these sources.

3) AGB stars with mass-loss
large enough to modify ($J-K_S$)$_0$ with respect to the isochrone by significant circumstellar dust
absorption ($\dot{M}$ $>\sim$10$^{-6}$ M$_{\odot}$/yr). A few such sources may be easily
identified from $K_S$-[15] (see Sect. \ref{sec_massloss} and Table \ref{tab_cat}) but disentangling their interstellar and circumstellar extinction remains difficult which adds uncertainty to their distance and luminosity determination.

A special case occurs for the most distant ISOGAL stars (mostly AGB) detected in $J$ and $K_S$ but
for which the isochrone would cut the red clump locus beyond the range it is detected (Fig. \ref{jjk_dist}).
However, one may still infer an approximate value of their distance by noticing that the isochrones
actually cutting the RCL, have approximately constant $\delta$($J-K_S$)/$\delta d$ spacing. One may thus extrapolate this to draw isochrones at larger distances (Fig. \ref{jjk_dist}).

Keeping these difficulties in mind, we give in Table \ref{tab_cat} an estimate of the reddening and
distance to the ISOGAL sources with detections in $J$ and $K_S$, assuming that they are red giants fitting the standard isochrone of \cite{Bertelli1994}, except for those having very large mass-loss as discussed in Sect. \ref{sec_massloss}, and those identified as young stars (see Sect. \ref{sec_yso}) as well as the foreground main sequence.
It is easy to infer the luminosity L$_{bol}$ from $K_{S0}$ with the well established bolometric correction, BC$_K$, for such red giant stars  close to 3 (see e.g. \citep{Frogel1987}) and Fig. 6 of \citet{Ojha2003}.

\subsection{Variations of extinction across the field and correlation with \element{C}\element{O} }

	As discussed above (Sect. \ref{distdeterm}), the extinction is not uniformly proportional to
the distance in the line of sight. The total extinction may be attributed to two main components - the first  being proportional to the average interarm  column density of dust. The second component of extinction is localized and associated with the excess of interstellar molecular gas in the spiral arms.

The effect of extinction in the Scutum-Crux arm should be responsible for the irregular patterns seen in the distribution of red clump stars along the red clump locus in the individual plots of each cell in Fig. \ref{jjk_dist} (see also the online material). The absence of stars in some sections of these loci should correspond to localized molecular clouds, with a jump in $J-K_S$ and thus in $A_\mathrm{V}$ on a very short distance where the number of red clump stars is very small; on the other hand, the sections with a regular large density of stars should correspond to regular smooth increase of $A_\mathrm{V}$.

\subsection{Extinction law at 7 and 15 $\mu$m and in the GLIMPSE bands}

\begin{figure*}
\includegraphics[width=0.32\textwidth]{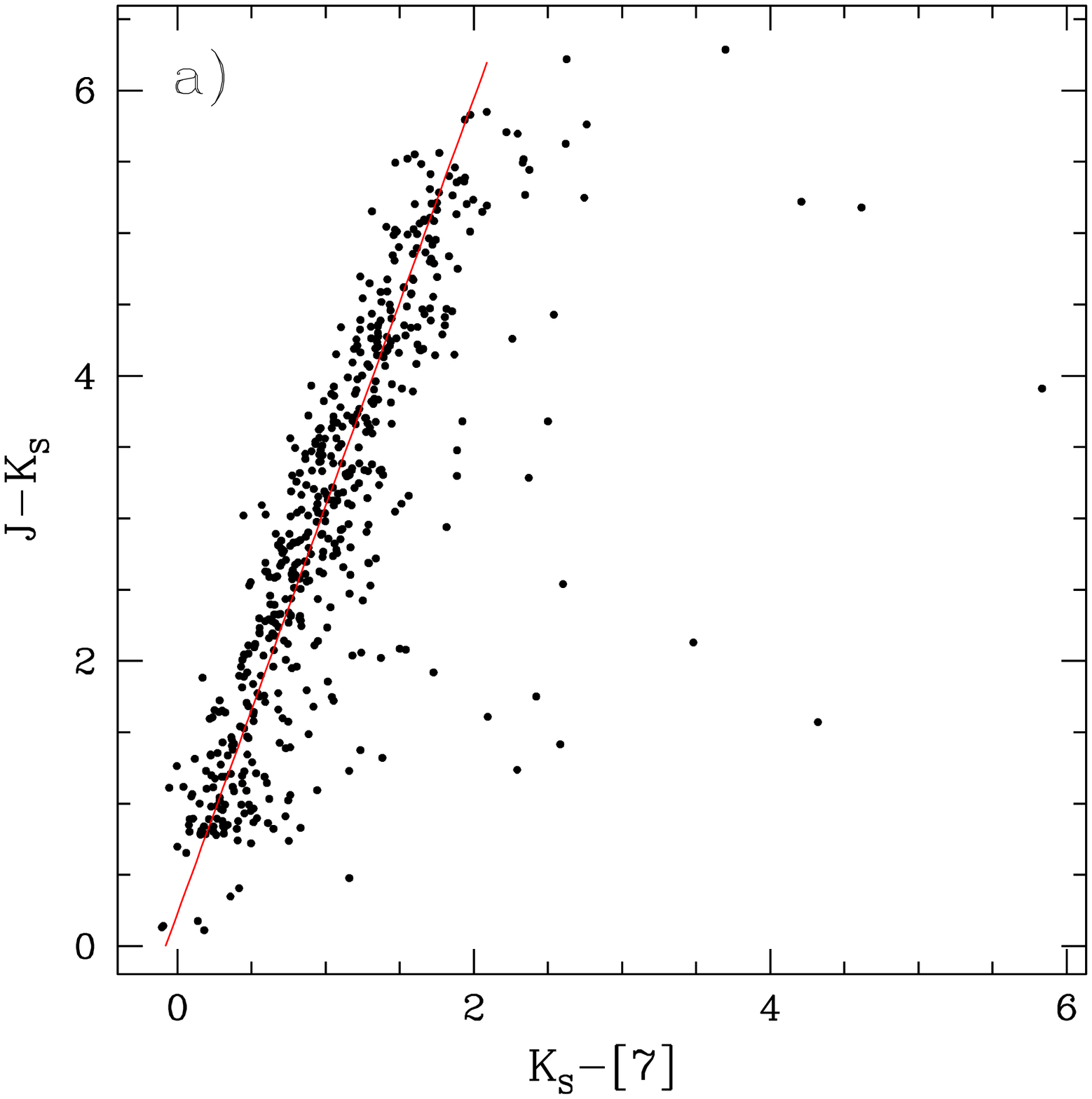}
\includegraphics[width=0.32\textwidth]{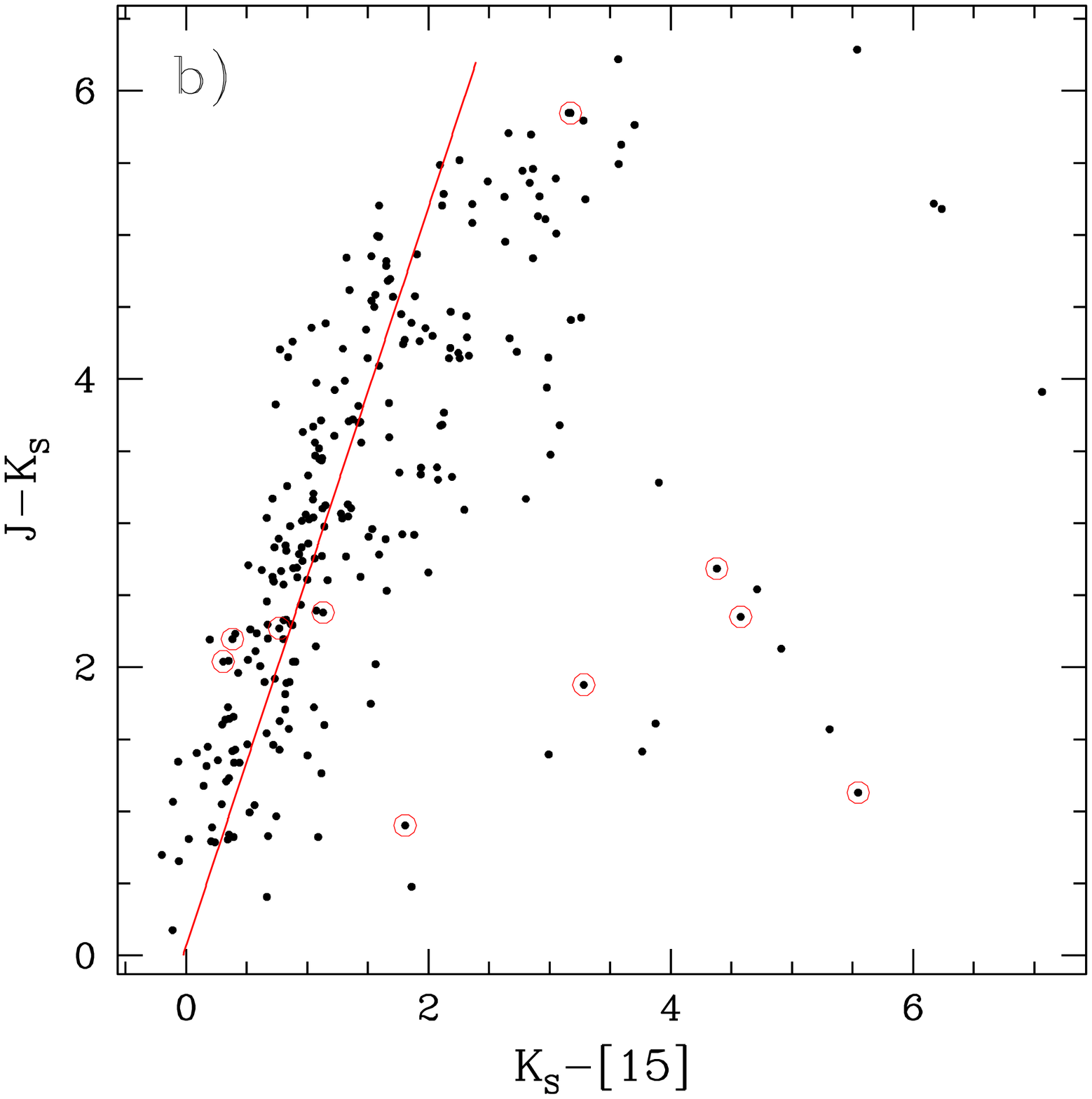}
\includegraphics[width=0.32\textwidth]{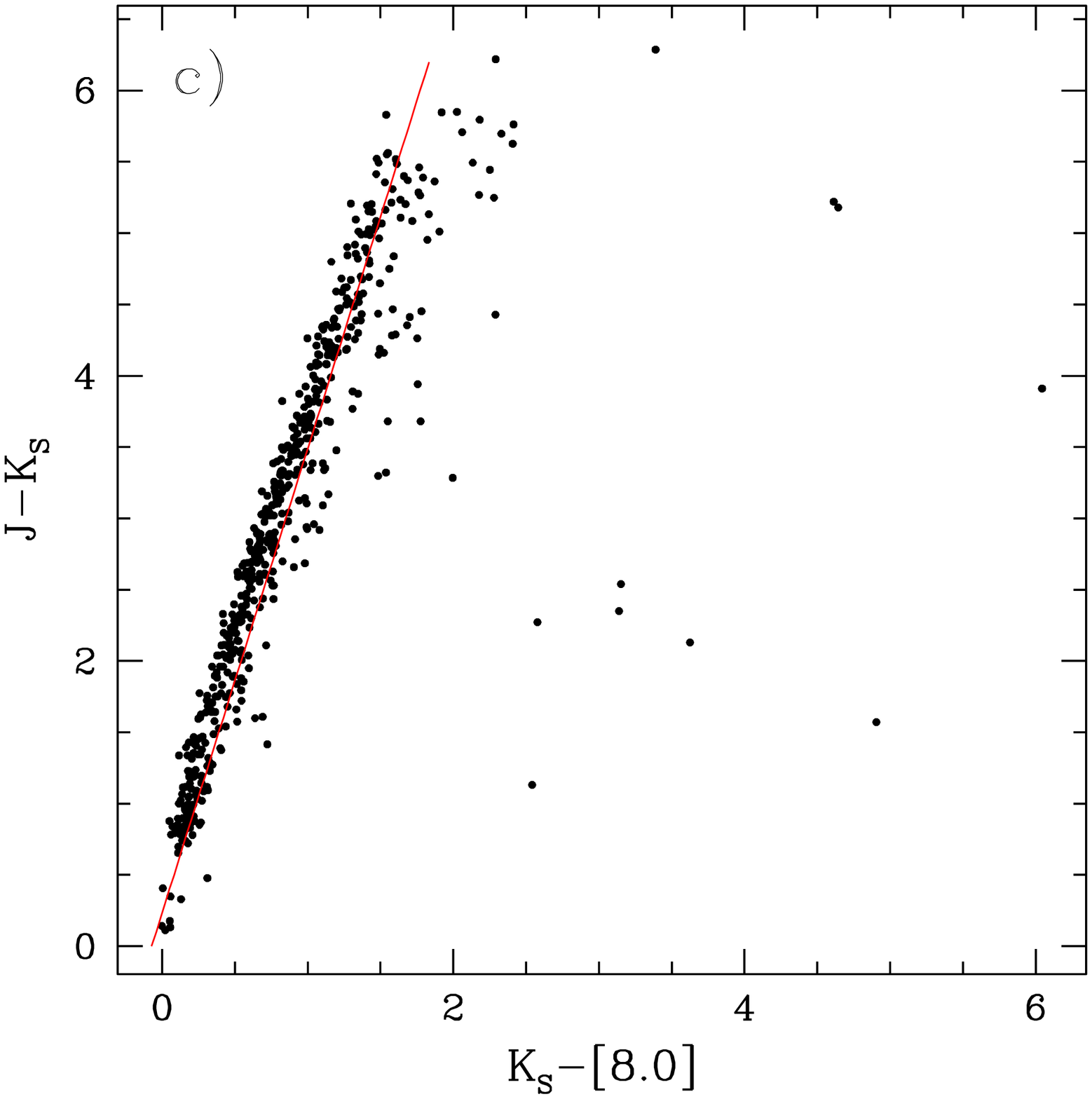}
\caption{(a) $J-K_S$ vs $K_S-[7]$ CCD showing the fit used to determine ($A_\mathrm{Ks}$-$A_\mathrm{[7]}$)/($A_\mathrm{J}$-$A_\mathrm{Ks}$) (b) $J-K_S$ vs $K_S-[15]$ CCD showing the fit used to determine ($A_\mathrm{Ks}$-A$_\mathrm{[15]}$)/($A_\mathrm{J}$-$A_\mathrm{Ks}$).  Sources not seen at [7] are shown with addtional open circles. (c) $J-K_S$ vs $K_S-\mathrm{[8.0]}$ CCD showing the fit used to determine ($A_\mathrm{Ks}$-$A_\mathrm{[8.0]}$)/($A_\mathrm{J}$-$A_\mathrm{Ks}$].  Only sources with ISO counterparts are shown here.
}
\label{jk_k7_fit}
\end{figure*}

	\citet{Jiang2003,Jiang2006} have estimated the interstellar infrared  extinction for the ISOGAL
fields based on the assumption that most of the  sources detected by ISOGAL are luminous RGB  or AGB
stars with moderate mass loss. Another assumption is that the intrinsic colours $(J-K_S)_\mathrm{0}$
 and  $(K-\mathrm{[7]})_0$ are similar for RGB and AGB stars, except for the sources with high mass loss.
\citet{Jiang2003} have discussed this point in detail. In $(J-K_S)$~vs.~$(K_S-\mathrm{[7]})$ colour-colour diagrams (CCD) most
of the  sources then follow  a straight line with some dispersion. The fitting of this straight line
thus determines the ratio [$A_\mathrm{Ks}$-$A_\mathrm{[7]}$]/[$A_\mathrm{J}$-$A_\mathrm{Ks}$]. We have repeated independently
the fitting of \citet{Jiang2006} for the LN45 field (Fig \ref{jk_k7_fit}a) and found
($A_\mathrm{Ks}-A_\mathrm{[7]}$)/($A_\mathrm{J}-A_\mathrm{Ks}$) = $0.35 \pm 0.02$, in agreement with their value.  Similarly, we have
found ($A_\mathrm{Ks}-A_\mathrm{[15]}$)/($A_\mathrm{J}-A_\mathrm{Ks}$) $\sim$0.39 (see Fig \ref{jk_k7_fit}b), with a larger uncertainty because of the increased
effect of mass-loss on ($K_S$-[15])$_\mathrm{0}$ (see below and \citet{Jiang2003,Jiang2006}).   Fig \ref{jk_k7_fit}c displays the $(J-K_S)$~vs~$(K_S-[8.0])$ figure for the GLIMPSE $IRAC4$ band.   A similar study has been done by \citet{Indebetouw2005} and they present extinction in the GLIMPSE IRAC bands ($A_\mathrm{IRAC}$) using similar diagrams towards a different GLIMPSE field in the galactic disk.  The values derived for ($A_\mathrm{Ks}$-$A_\lambda$)/($A_\mathrm{J}$-$A_\mathrm{Ks}$) for the LN45 field are tabulated in Table \ref{tab_akl_ajk} and are consistent with \citet{Jiang2006} and \citet{Indebetouw2005}.  Note however, that different extinction coefficients have been derived by various authors depending on the line of sight.  For example, \citet{Flaherty2007} derived different extinction coefficients over a large range of near and mid-infrared wavelengths towards nearby star-forming regions.   They also obtained slightly different numbers from the data analysed by \citet{Indebetouw2005}. \citet{Flaherty2007} obtained $A_\mathrm{[24]}/A_\mathrm{V}$ values of  0.039 to 0.046 towards two star forming regions and these are larger than the value we have obtained for $A_\mathrm{[15]}/A_\mathrm{V}$ (=0.024).   \citet{RomanZuniga2007} found larger extinction coefficients for the IRAC filters towards a dark cloud core.

\begin{table}
\caption{$A_\lambda / A_\mathrm{V}$ for different filters}
\label{tab_akl_ajk}
\begin{tabular}{ccccl}
\hline
Filters&  $\lambda_{c}$ & $\Delta \lambda$ &$A_\lambda / A_\mathrm{V}$ & Notes\\
\hline
$J$&1.25&0.25&0.256&2MASS\\
$H$&1.65&0.3&0.142&2MASS\\
$K_S$&2.17&0.32&0.089&2MASS\\
$LW2$  &  6.7&3.5  & 0.031&ISOGAL\\
$LW3$  & 14.3& 6 & 0.024&ISOGAL\\
$IRAC1$&3.58 & 0.75 &0.046&GLIMPSE\\
$IRAC2$&4.50& 1.02 &0.042&GLIMPSE\\
$IRAC3$&5.80 & 1.43 &0.036&GLIMPSE\\
$IRAC4$&8.00 & 2.91 &0.037&GLIMPSE\\
\hline
\end{tabular}
\end{table}

\section{Nature of sources} 
\label{sec_nature}

\begin{figure*}
\includegraphics[width=0.67\textwidth]{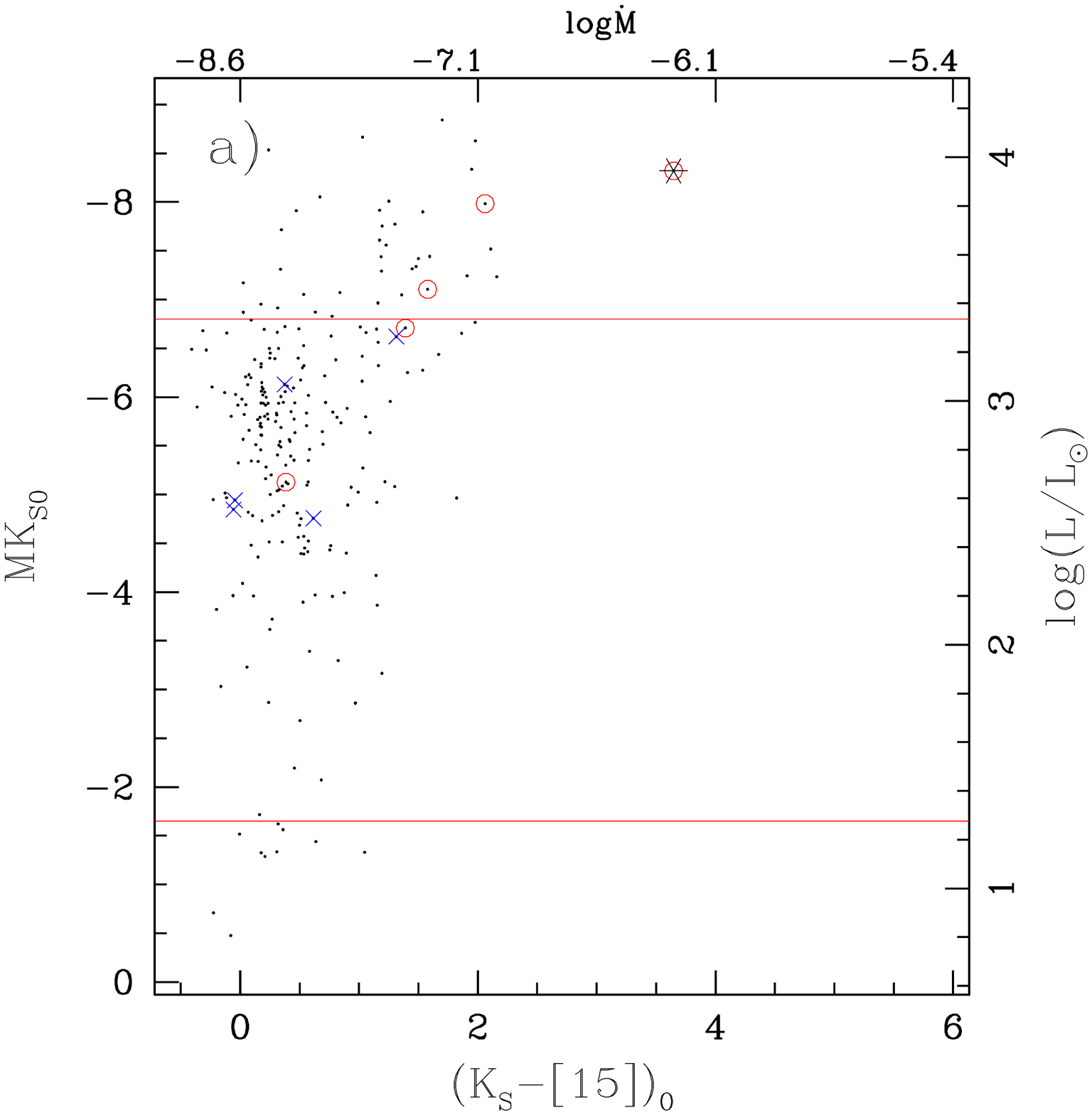}
\includegraphics[width=0.67\textwidth]{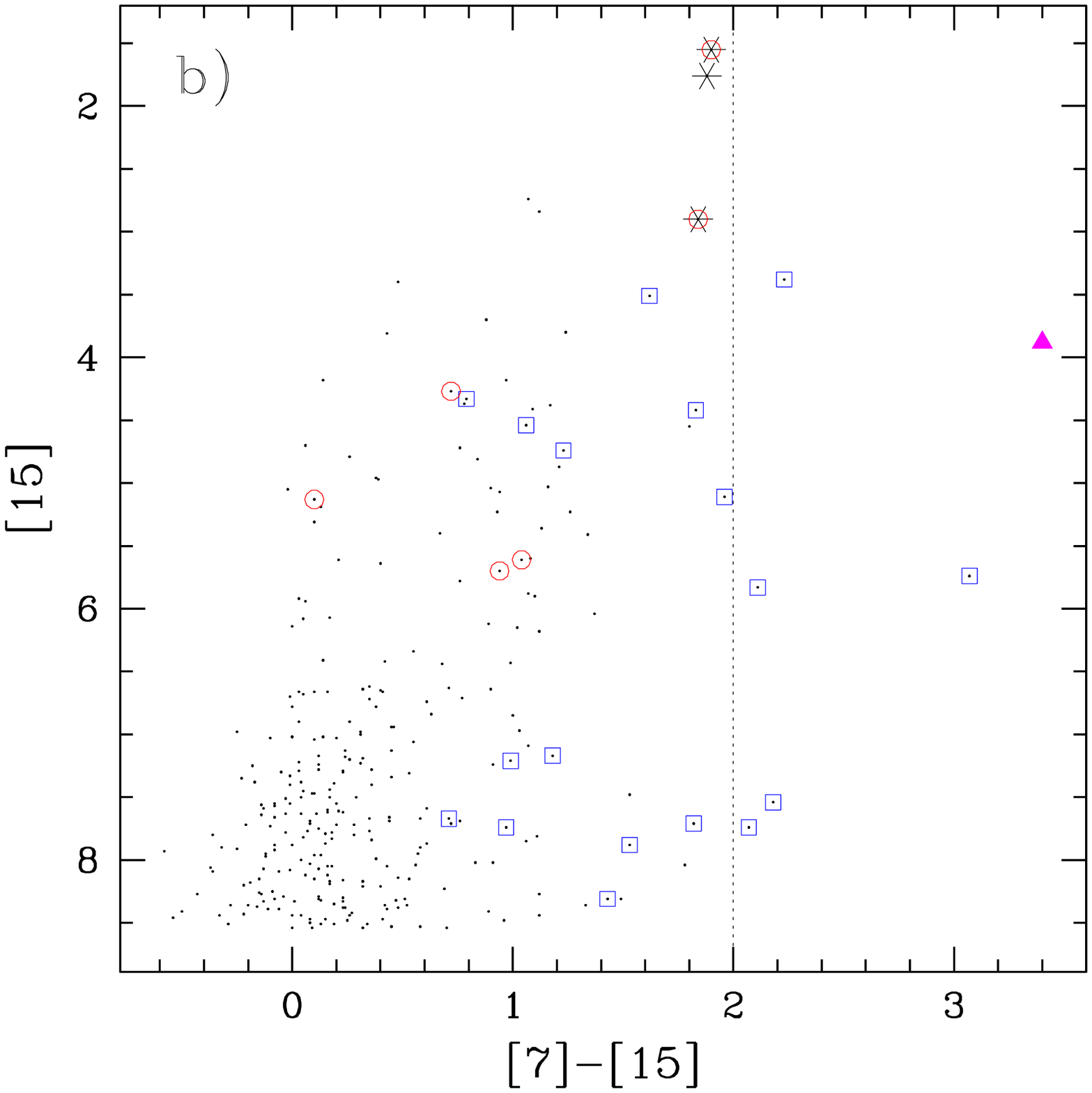}
\caption{(a) $\mathrm{M}K_{S0}$ vs $(K_{S}-\mathrm{[15]})_\mathrm{0}$ diagram.  Horizontal lines at
$\mathrm{M}K_{S0}$ of $-6.8$ and $-1.65$ show the RGB tip and the red clump magnitudes
respectively.  Variables in [15] are marked by asterisks.  Variables in $K_S$ are represented by
open (red) circles.  Non-detections at [7] have additional blue crosses.  Mass-loss rate
($\log\dot{M}$) are marked for corresponding colours on the top axis.  See Sect.
\ref{sec_massloss}.  Luminosities are labelled on the right axis.  YSO candidates are not
shown (b) The ISOGAL [7]-[15] vs [15] CMD showing the various types of sources: variables (15$\mu$m, $K_S$), YSO, planetary nebula (magenta triangle, see text).
YSO candidates are marked with blue open squares.  All other symbols are as in (a).}
\label{Mk_k015}
\end{figure*}

Figure \ref{Mk_k015}a shows the $(K_S-\mathrm{[15]})_\mathrm{0}$ vs. $M_{Ks0}$ diagram for stars to which the distance has been derived.  Note that this implies that these sources have $J$ and $K_S$ counterparts. Theoretical tip of the RGB as well as the position of the red clump provided by the isochrones are also indicated. The bulk of sources are RGB stars while a few K giants are also present in our sample.

One interesting source (number 729 in the LN45 catalog) has an extremely large $(K_S-[15])_\mathrm{0} > 8$. It is very bright at [7] and  at [15].   In the [7] vs. $K_S$-[7] diagram (Fig. \ref{ysofig}c) this object is situated at a location ([7]=3.64, $K_S$-[7]=7.38) populated by high mass-losing AGB stars \citep{Schultheis2000}. The derived distance of 5~kpc for this object is certainly too low (due to uncertain 2MASS $J$ band photometry) which means that the absolute magnitude must be much brighter.  This source is not shown in the Fig. \ref{Mk_k015}a due to its distance being very uncertain.

ISOGAL [7]-[15] vs [15] CMD is shown in Fig. \ref{Mk_k015}b.  The various stellar populations detected by ISOGAL
are plotted with different symbols in this figure.  This figure is further discussed in the Sect. \ref{sec_yso}.

ISOGAL-P J143318.1-604938 (number 636 in the LN45 catalog) is actually the well known planetary nebula ESO 134-7 \citep{Kerber2003}.  This planetary nebula is notable for it's exceptionally high velocity wind and [WN] nucleus \citep{Morgan2003}.  This has been miss-identified as a YSO candidate by \citet{Felli2002}.  This source is marked by a magenta triangle in Fig. \ref{Mk_k015}b and does not appear in Fig. \ref{Mk_k015}a due to uncertain distance.

\subsection{AGB stars and mass-loss}
\label{sec_massloss}
\subsubsection{Identification, distance, variability, dereddening and luminosity of AGB stars}

\begin{figure*}
\includegraphics[width=0.67\textwidth]{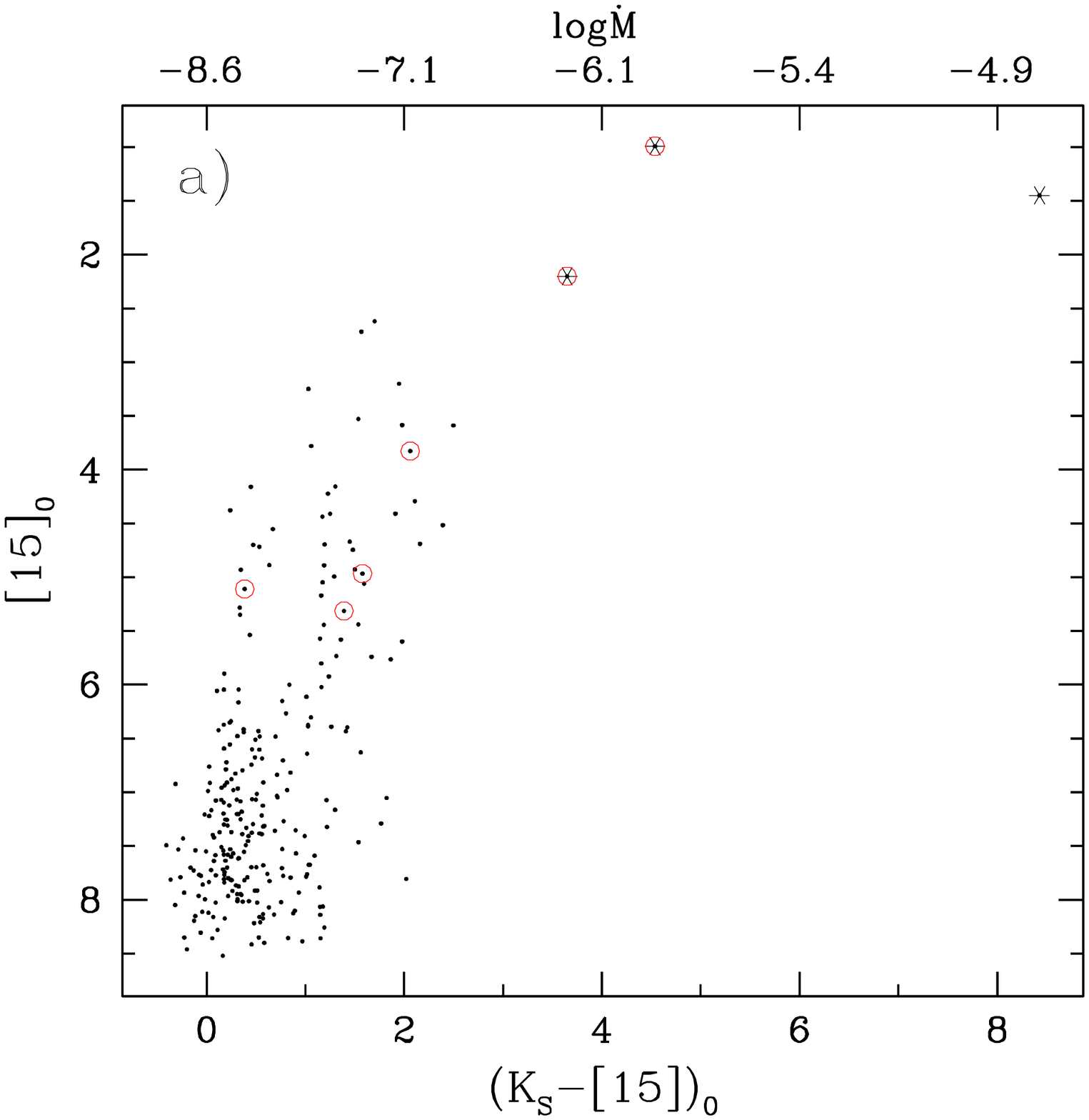}
\includegraphics[width=0.67\textwidth]{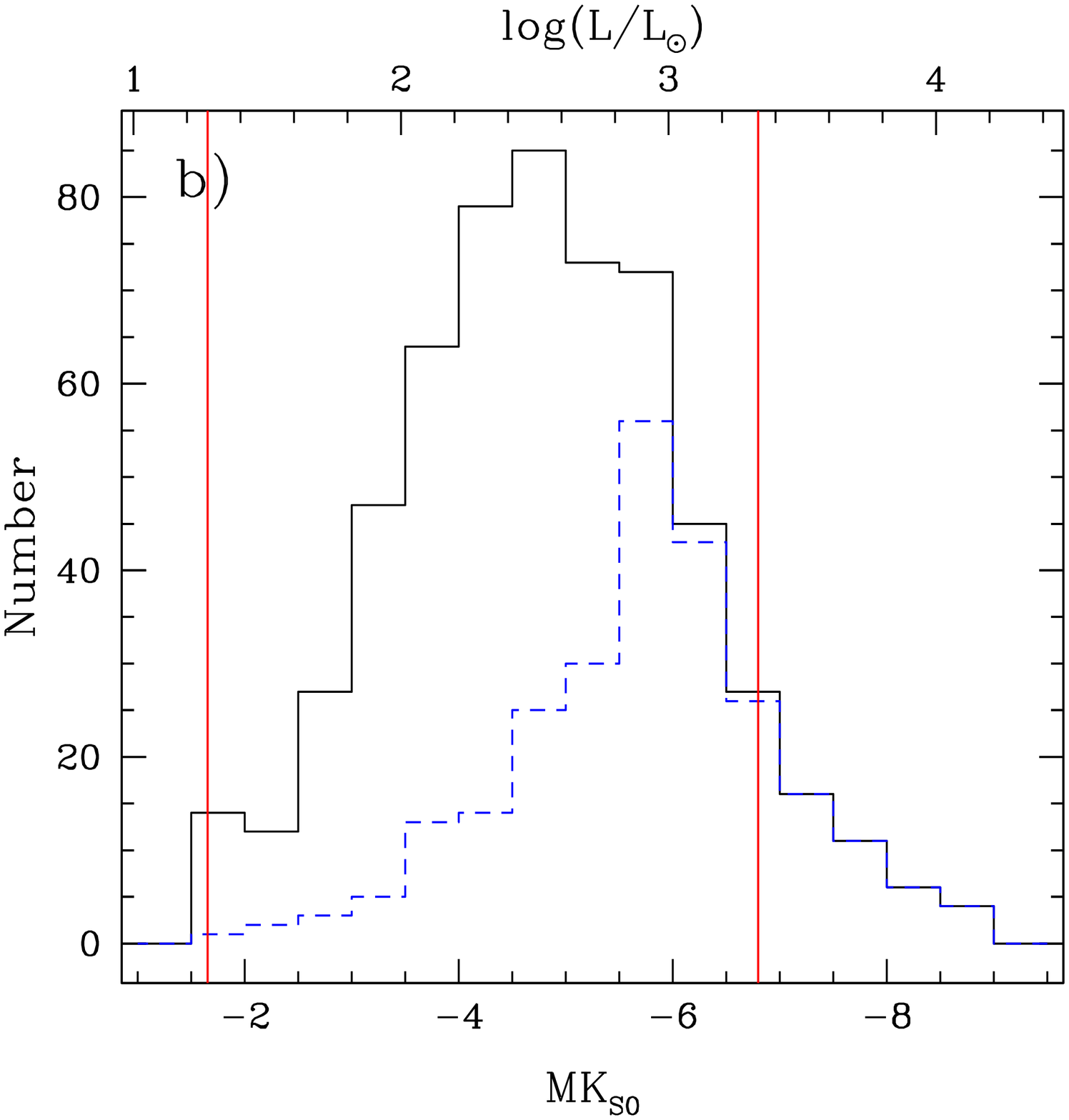}
\caption{ (a) The $(K_S-\mathrm{[15]})_\mathrm{0}$ vs [15] diagram.  Candidate variable stars from the near-infrared photometry (see text) are shown as (red) open  circles. Asterisks are sources with  more than 0.5mag difference in the two [15] observations.  The top axis is labeled with values of $\log\dot{M}$. We have included the 2MASS photometry for the sources saturated in GLIMPSE.  Note that YSOs and the known planetary nebula are not included in the plot.  (b) Histogram of  $\mathrm{M}K_{S}$ in bins of 0.5mag for sources brighter than the red clump ($\mathrm{M}K_{S0}=-1.65$).    Solid line shows the histogram for all ISOGAL sources with 2MASS counterparts (including sources saturated in GLIMPSE and those with $J$ band missing in GLIMPSE--2MASS).
Dashed (blue) line is the histogram for the sources which have an ISOGAL 15$\mu$m counterpart. YSOs and the known planetary nebula are not included. The red clump and the tip of the RGB are shown as solid vertical lines at $\mathrm{M}K_{S0}$ of -1.65 and -6.8 respectively.
}
\label{k015_15}
\end{figure*}

Fig. \ref{k015_15}a shows the $(K_S-\mathrm{[15]})_\mathrm{0}$ vs. $\mathrm{[15]_0}$ diagram of the LN45 field. Only those sources believed to be AGB or RGB stars
with derived distances are plotted, so that young stellar objects are not included. As
discussed by \citet{Omont1999} and \citet{Glass1999} this diagram represents a mass-loss sequence.  We have superimposed objects which are variable in $K_S$ and $J$ (based on the 2MASS and the DENIS data which were observed at different epochs) with an amplitude larger than 0.4 mag in $J$ and 0.2 mag in $K_S$.  We found in total 5 variable AGB stars which populate the typical region of long period variables (LPV) \citep[see also][]{Glass1999} and one object  (number 551 in Table \ref{tab_cat}) with an absolute magnitude $\mathrm{M}K_{S} = -5$  below the RGB tip.  Its colour in $(K_S-\mathrm{[15]})_\mathrm{0}$ suggests a red giant classification and as it has small $J-K_S$ colour, the star is expected to be nearby.  A total of 3 sources (numbers 233, 424 and 729 in Table \ref{tab_cat}) were found to be variable (amplitude > 0.5 mag) in the repeated 15$\mu$m observations.   Note that this variable identification has been inferred from observations of only two
epochs. Therefore, the actual number of LPVs is certainly much larger, particularly the sources with luminosities larger than the RGB tip. \citet{Schultheis2000} claim that about 40\% of the variable stars can be recovered using only two epoch measurements.

Figure \ref{k015_15}b shows the luminosity function in the $\mathrm{M}K_{S}$ band for the RGB
and AGB stars. The tip of the RGB and the red clump are indicated. The
peak in the general distribution is at $\mathrm{M}K_{S}=-5$, while those sources
with 15~$\mu$m ISOGAL counterpart peak at one magnitude brighter.
This is due to the reduced sensitivity and completeness of the
15~$\mu$m flux. We detect all  AGB stars at 15~$\mu$m in this
direction, while the 15~$\mu$m sources  get incomplete just below the
RGB-tip.

\subsubsection{Mass-loss rate determination from 15$\mu$m excess}

It is generally agreed that the mid-infrared excess ([$\lambda_1$]-[$\lambda_2$])$_\mathrm{0}$ is a
good indicator of the mass-loss rate $\dot{M}$ of AGB stars \citep{Whitelock1994,LeBertre1998,Ojha2003}.
\citet{Ojha2003} have shown that $(K_S-\mathrm{[15]})_\mathrm{0}$ provides a good estimate of
$\dot{M}$, and based on this they have determined $\dot{M}$ of AGB stars in the ISOGAL fields of the intermediate bulge (see
also \citet{Alard2001}). \citet{Ojha2003} discussed two models to relate AGB infrared colors to
mass-loss rates.
We have used dust radiative transfer models by \citet{Groenewegen2006} for oxygen-rich AGB stars.
The mass-loss rates are based on a model for an oxygen-rich AGB star with $\mathrm{T_{eff}}$ of
2500~K and  100\% silicate composition. From the set of values of $\dot{M}$ computed by \citet{Groenewegen2006}, we have used the following relation between $(K_S-\mathrm{[15]})_\mathrm{0}$  and the mass-loss rate $\dot{M}$
\begin{figure}[h]
\includegraphics[width=\columnwidth]{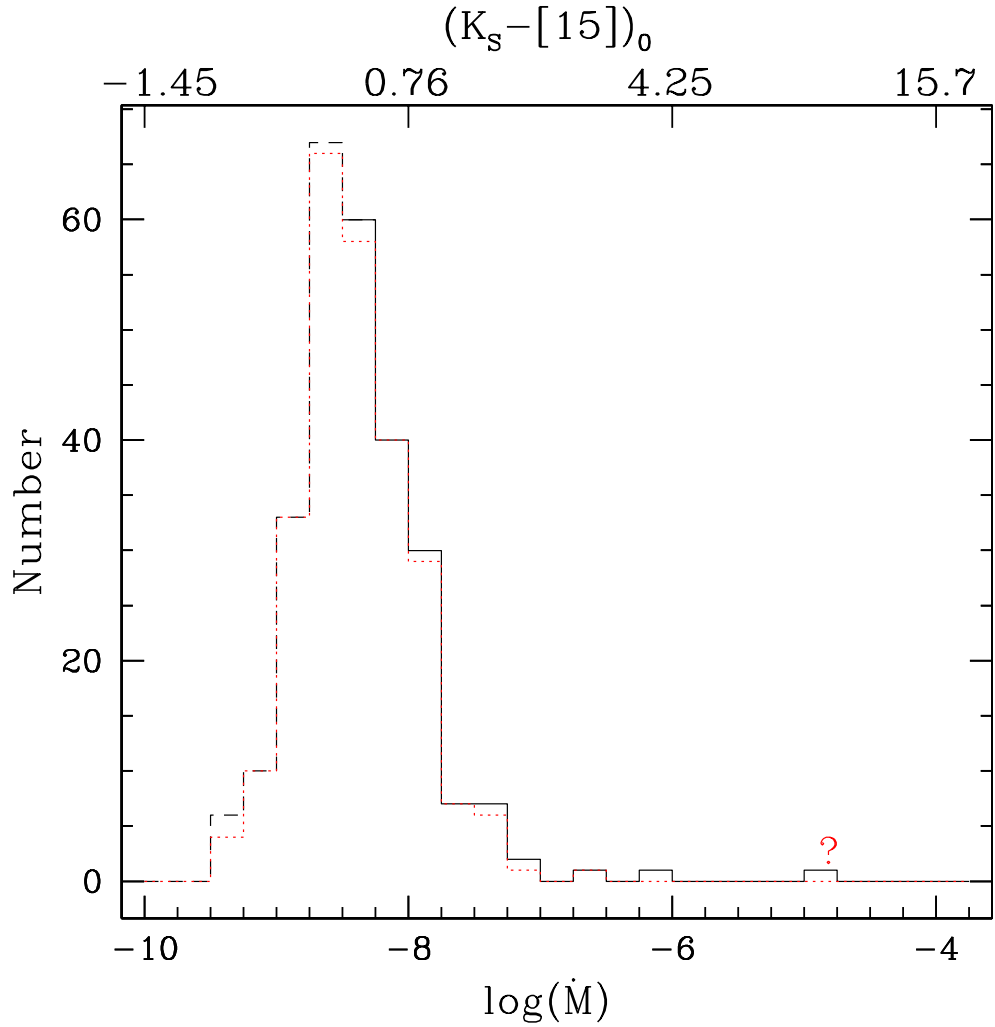}
\caption{Histogram of the mass loss rates (bottom axis) determined from the $(K_S-\mathrm{[15]})_\mathrm{0}$ colours (top axis). The (red) dotted histogram is obtained after discarding sources with $J$ band missing in the GLIMPSE--2MASS catalog. Histogram for the mass loss rates below the value of $\log \dot{M}$ = -8 are represented by dashed line.  The contribution of source 729 with uncertain distance (see text)  is marked by a question mark.}
\label{mdothist}
\end{figure}

\begin{equation}
	\log\dot{M}~=~-8.6171~+~0.85562~X~-0.064143~X^2~+~0.0018083X^3
\end{equation}
where $\dot{M}$ is in M$_{\odot}$ yr$^{-1}$ and $X$ is $(K_S-\mathrm{[15]})_\mathrm{0}$.  We have derived mass-loss rates for ISOGAL sources of LN45 field from this relation and the mass-loss rates are listed in Table \ref{tab_cat}. Note that this relation is valid for $\mathrm{-10.0 < log\dot{M} < -4.7}$, but the accuracy of our data limits its  use to  $\dot{M}$~$ > $~10$^{-8}$M$_\odot$ yr$^{-1}$.

Figure \ref{mdothist} shows the histogram of the derived values of the mass-loss rates of the AGB stars in the LN45 field. Most of the sources are RGB stars with $(K_S-\mathrm{[15]})_\mathrm{0}$ $<$ 1 (i.e. $\log \dot{M} < -7.8$).
\citet{Jiang2003} studied a similar field at $\ell=-18 \deg$ (FC-01863+00035). Their $(K_S-\mathrm{[15]})_\mathrm{0}$ vs [15] diagram look similar to ours.

In most cases, the value of $(K_S-\mathrm{[15]})_\mathrm{0}$ is derived from the observed $K_S$ and 15$\mu$m magnitudes and by applying the reddening values previously determined from the red giant isochrone.  However, one
should note that an appreciable error in the value of $(K_S-\mathrm{[15]})_\mathrm{0}$ may result for stars with large mass loss rates as $K_S$ and 15$\mu$m observations are from different epochs and such stars are strong variables.

   The overall observational uncertainty is rather large for the derived values of
$\dot{M}$ for individual stars. The combined errors in photometry and reddening
determination and the effect of variability  result in a global rms uncertainty of
$(K_S-\mathrm{[15]})_\mathrm{0}$ in the range 0.3--0.5 mag.  The variation of (K$_s$-[15])$_0$ leads to an   uncertainty in  $\dot {M}$ which could be up to a factor of 3 for $\dot {M}$ $\sim$ 10$^{-7}$ M$_{\odot}$/yr, which corresponds to most of the sources with large mass-loss (Fig. \ref{mdothist} see Fig. \ref{k015_15}a).

Alternatively, one could have also inferred $\dot{M}$ from ($K_S-\mathrm{[7]})_\mathrm{0}$ or from [7]-[15] colours.  Deriving $\dot{M}$ from [7]-[15] \citep{Alard2001} has the advantage that the latter
quantity depends very little on the extinction. However, this method is less sensitive since
the variation in [7]-[15] is small. $(K_S-\mathrm{[7]})_\mathrm{0}$ has  a low sensitivity for small mass-loss rates, but could be conveniently used for high mass-loss rates.

\subsection{Candidate young stellar objects}
\label{sec_yso}

\begin{figure*}[h]
\includegraphics[width=0.52\textwidth]{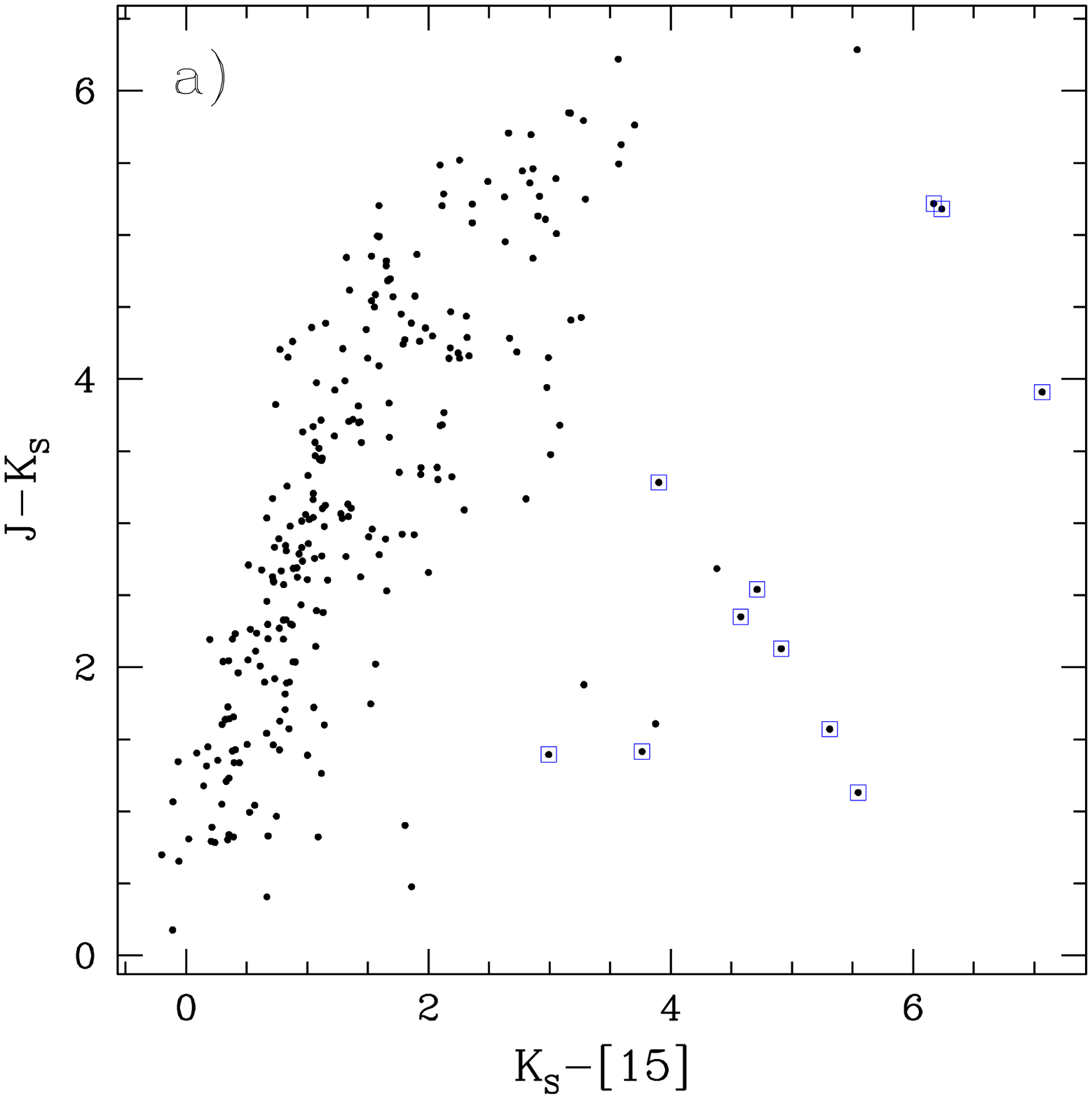}
\includegraphics[width=0.52\textwidth]{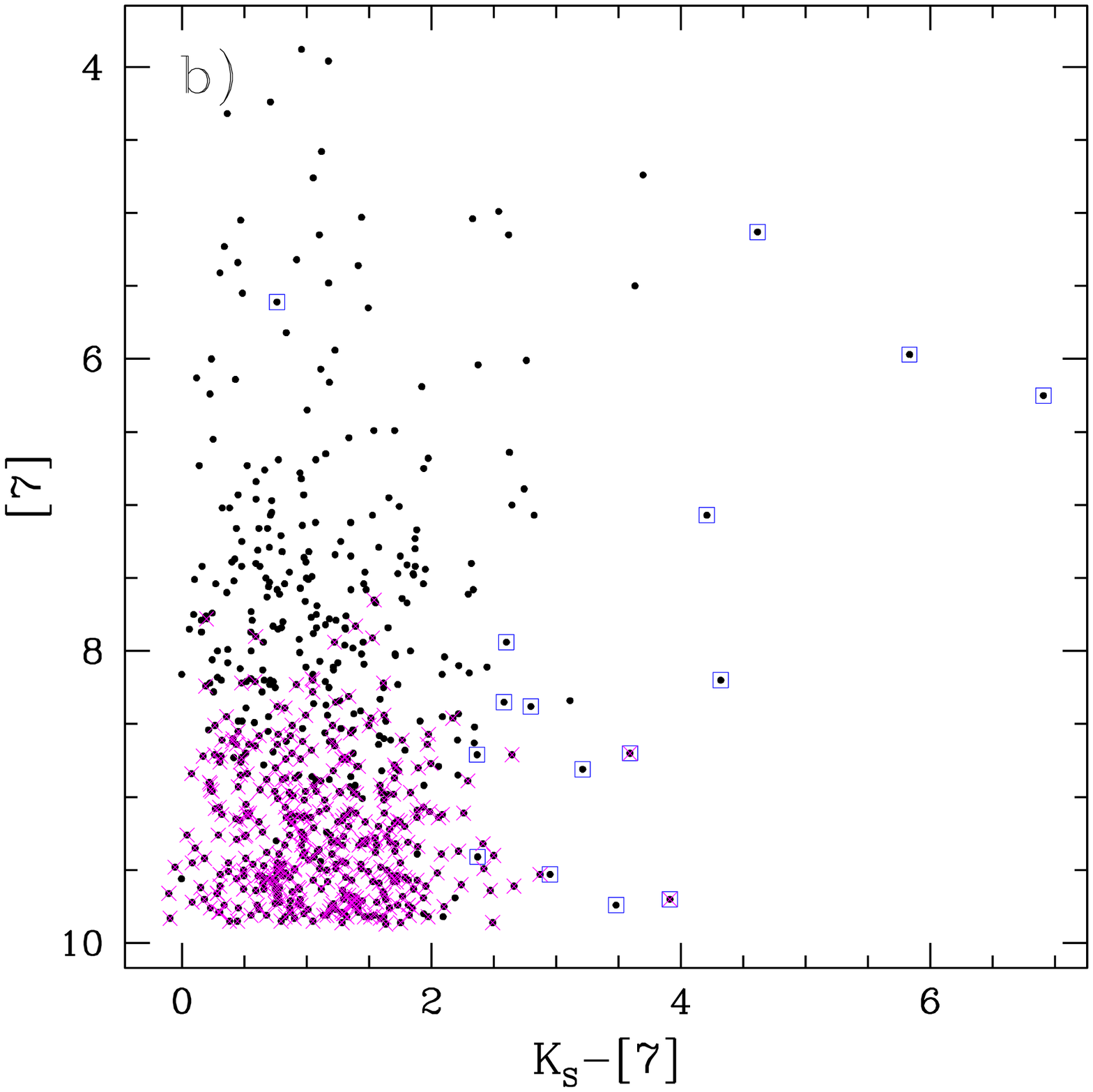}
\caption{(a) $J-K_S$ vs $K_S$-[15] CCD with the YSO candidates plotted as open (blue) squares.   (b) [7] vs $K_S$-[7] CMD.   Magenta crosses are sources not detected at [15].}
\label{ysofig}
\end{figure*}

\begin{figure*}[h]
\includegraphics[width=0.52\textwidth]{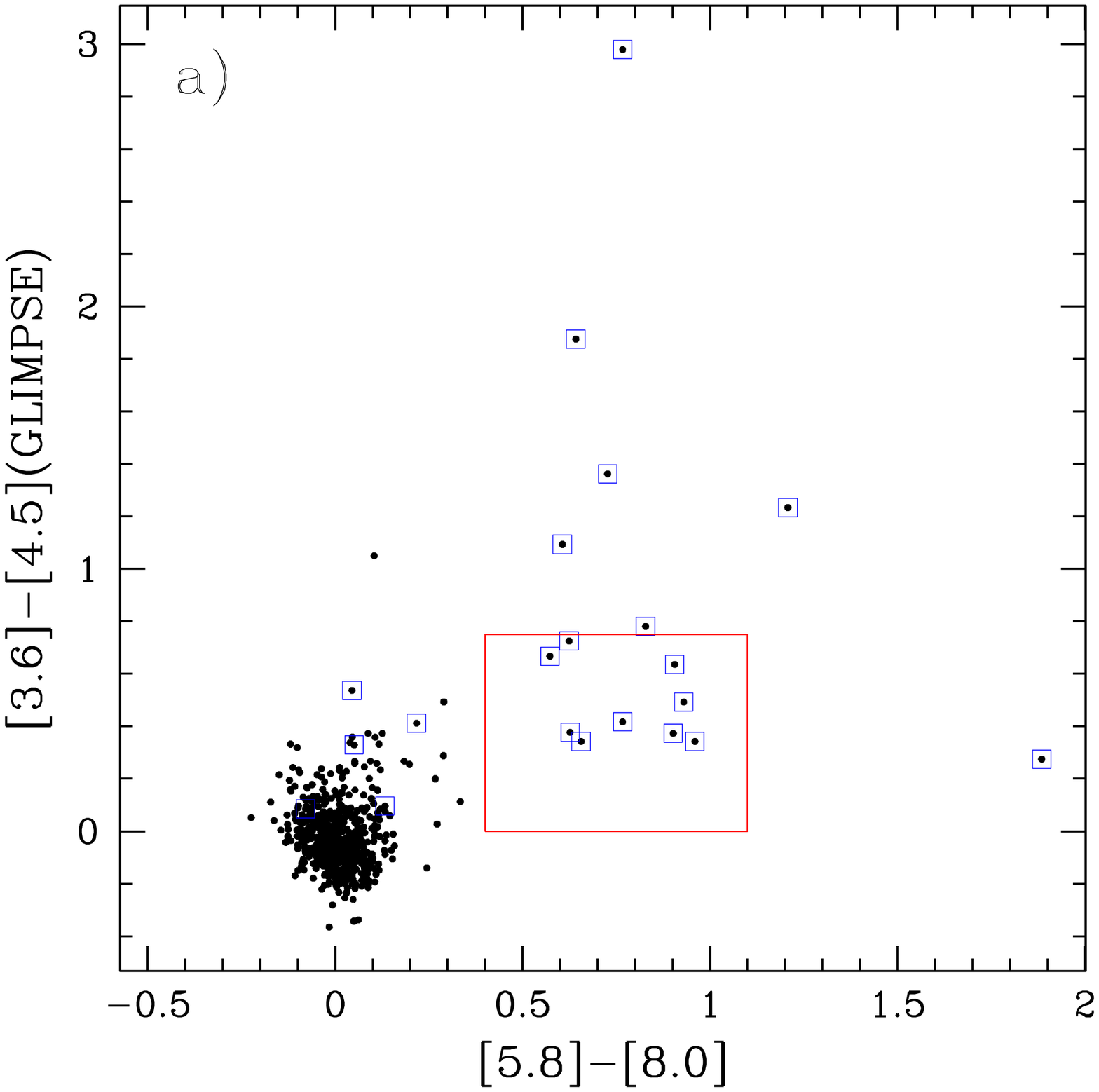}
\includegraphics[width=0.52\textwidth]{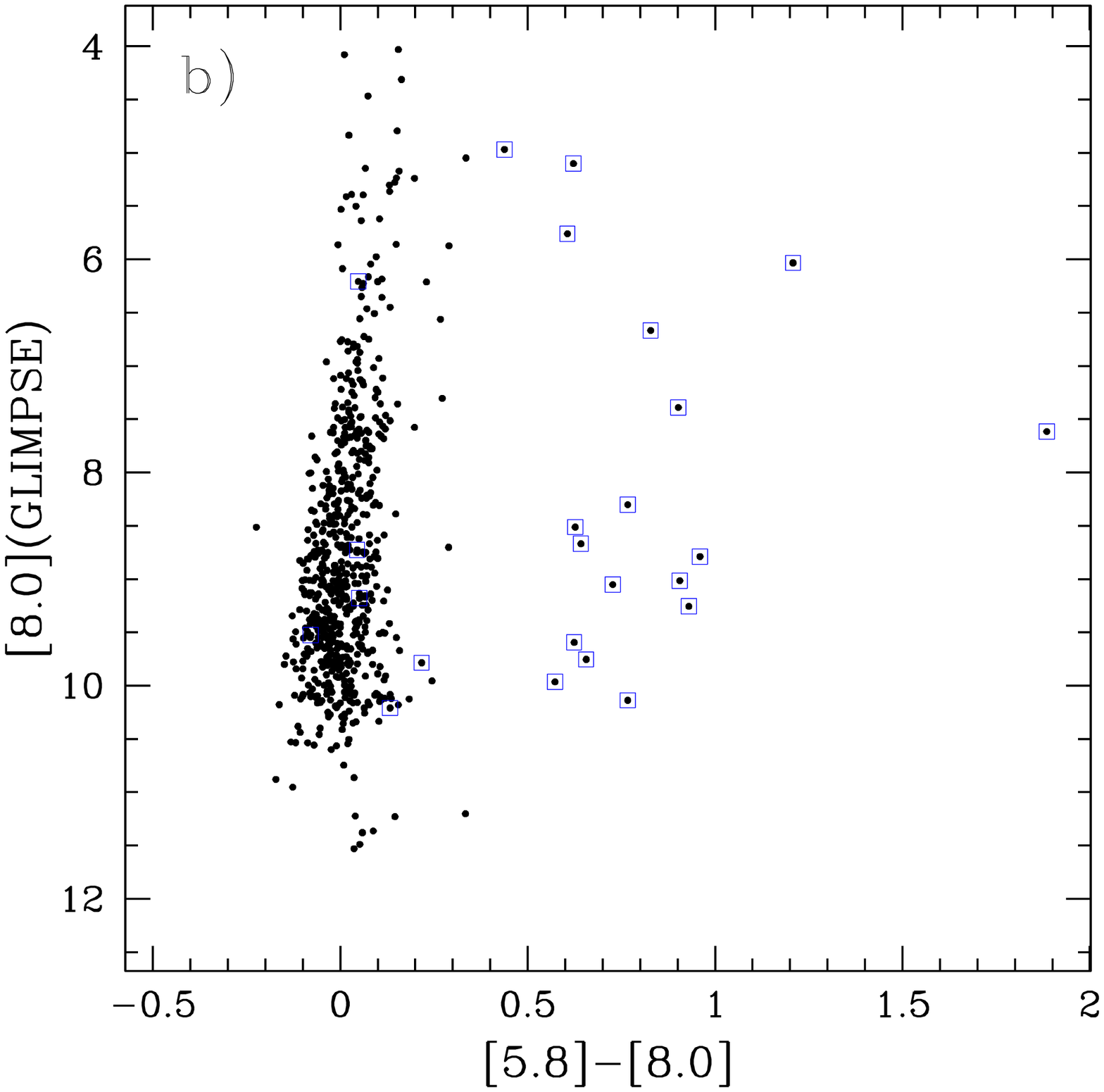}
\caption{(a) GLIMPSE [5.8]-[8.0] vs [3.6]-[4.5] CCD. The class II YSO region of \citet{Allen2004} is demarcated by a red box.  (b) GLIMPSE [5.8]-[8.0] vs [8.0] CMD. YSO candidates are shown as open (blue) squares in both panels.}
\label{ysofig2}
\end{figure*}

In the current work, sources with an excess in the mid-infrared  (see Fig. \ref{ysofig}a) and relatively small near-infrared colours, ($J-K_S$) are identified as Young Stellar Object candidates (YSO).  The properties listed above are indicative of relatively nearby, less extincted sources so that the mid-infrared excess has to arise from circumstellar dust in which the star may  be embedded.    All of these 15$\mu$m sources have 7$\mu$m counterparts and the 7$\mu$m also shows an excess in the $K_S-\mathrm{[7]}$ colour (see Fig. \ref{ysofig}b).  There are two faint 7$\mu$m sources without 15$\mu$m counterparts.   We include these in the list of candidate YSOs as they have GLIMPSE colours (Fig. \ref{ysofig2}a) consistent with those expected for YSOs.

In the past, the characteristic ISOGAL diagram, [15] vs [7]-[15] (Fig. \ref{Mk_k015}b) has been used by several authors \citep[e.g.][]{Felli2000,Felli2002,Schuller2006} to identify candidate YSOs.   \cite{Felli2002} identified a simple criteria ([7]-[15]$~>~2.0$) to qualify a source as a candidate YSO.  In LN45 field, we find six sources which satisfy the \cite{Felli2002} colour criteria.  However, one of them is a well known planetary nebula (ESO 134-7) as mentioned earlier.
Most of the sources with [7]-[15]$~>~2.0$ are relatively fainter than the limit required by \cite{Felli2002}.
Note that this colour criterion is neither a sufficient nor a necessary condition as evinced by the large number of YSO candidates with relatively smaller colours.  \citet{Schultheis2003} have shown that the ISOGAL [7]-[15] colour, by itself, is not sufficient to identify YSO candidates and one requires additional data such as  spectroscopic followup.

Figures \ref{ysofig2}a and \ref{ysofig2}b show the GLIMPSE CCD and CMD.  In the GLIMPSE CCD we have marked the area occupied by class II YSOs \citep{Allen2004} by a rectangular box.  Ten 15$\mu$m sources identified as YSOs occupy this region.   Most of the other sources have colours consistent with class I types (i.e. with redder GLIMPSE colours).   A few  YSO candidates have relatively small GLIMPSE colours.  These are sources with an excess in the $K_S$-[15], [7]-[15] colours and are relatively faint but still significant at 15$\mu$m. Additional information would be needed to determine their true nature.

\section{Conclusion} 

We have studied a large ISOGAL field towards the Galactic disk and established a standard mode of studying the large set of ISOGAL observations of the disk.  We show that the  stars of  the  red clump can be traced all along the line of sight using the 2MASS $J$ and $K_S$ data up to the start of the  Scutum--Crux  arm  towards  the  $\ell=-45\degr$ direction.  We find  that the  locally accepted  value of  the extinction per unit distance, $c_\mathrm{J}$, is sufficient to  fit well the red  clump locus up to 2.5kpc in this direction.
However,  beyond 2.5kpc,  $c_\mathrm{J}$ varies  with  galactic  latitude  and increases with distance.  We use the red clump  locus to obtain the distance and  extinction towards  individual stars  assuming them  to be  red giants  and thus  following the  RGB/AGB isochrone.  The  distribution of  stellar  density  rises as one hits the spiral arm   at  4kpc.  The 2MASS  data are  not deep enough to  detect    the stars  of  the  red clump at distances larger than 4kpc at the lowest galactic latitudes.   We do see more luminous AGB  stars to  larger distances.

Most of the red giants brighter  than  the stars of the red  clump are detected by  ISOGAL at 7$\mu$m.  The ($K_S$-[15])$_0$ colour provides the mass loss  rates for the AGB stars.  There are not many AGBs  with large  mass-loss rate in  this  direction.

From the mid infrared colour excess we identify a total of 22 YSO candidates in this field.

We provide a catalog of the sources detected by ISOGAL with the estimated extinction and distance.  We tabulate also the mass-loss rates for the several hundred red giants towards this field.

\medskip
{\it Acknowledgements.}

This work is based on observations made with the {\em Spitzer Space Telescope},
which is operated by the Jet Propulsion Laboratory,
California Institute of Technology under a contract with NASA.
We are grateful to the GLIMPSE project for providing access to the data.
This publication makes use of data products from the Two Micron All Sky Survey, which is a joint project of the University of Massachusetts and the Infrared Processing and Analysis Center/California Institute of Technology, funded by the National Aeronautics and Space Administration and the National Science Foundation.
This research made use of data products from the Midcourse Space Experiment.  Processing of the data was funded by the Ballistic Missile Defense Organization with additional support from NASA Office of Space Science.  This research has made use of the SIMBAD database, operated at CDS, Strasbourg, France.
We acknowledge use of the TOPCAT software from Starlink and the VOPLOT virtual observatory software from VO-India (IUCAA) in this work.
We thank the referee, Sean  Carey, for constructive comments and criticisms which have improved the paper.
S. Ganesh was supported by Marie-Curie EARA fellowship to work at the Institut d'Astrophysique de Paris.
We are grateful to the Indo-French Astronomy Network for providing travel support to enable M. Schultheis to visit PRL.
Work at the Physical Research Laboratory is supported by the Department of Space, Govt. of India.




\Online
\begin{appendix}
\begin{onecolumn}
\section{Flag definition for ISOGAL--GLIMPSE association : {\it readme} file}
\label{flag_readme}
\begin{verbatim}
r0 = radius of association for 10% chance of false association
r0 = 3.8 for new GLIMPSE catalog with 4601 IRAC4 (i4) sources for LN45
r1 = 2.3'' and r2 = 1.5 for LN45 field
m7 - LW2 detection;  m15 - LW3 detection;
m81 - nearest GLIMPSE source ;
m82 - next (second) nearest GLIMPSE source


Flag = 9: Saturated in GLIMPSE i4 :
	ISOGAL (m7) brighter than 4 and no source in GLIMPSE i4
Flag = 8: Not at saturation level but no GLIMPSE - could be due to extendedness
	ISOGAL (m7 & m15) present and (m7 fainter than 4 and m15 brighter than 5) and
             no source in GLIMPSE i4

Flag = 7: Not at saturation level and GLIMPSE source in  r > r0 and r < 4.5''

Flag = 5:  very secure associations
  (separation < 1.5 && abs(m7-i4) < 0.3)
  i.e. r < r2 and |m7-m81| < 0.3 and no other GLIMPSE source within r0

Flag = 4:  secure association
(1.5 < separation && separation < 2.3 && abs(m7-i4) < 0.5)||
(separation < 1.5 && abs(m7-i4) < 0.5 && abs(m7-i4) > 0.3)
4.1 r2 < r < r1 and |m7-m81| < 0.5 and no other GLIMPSE source within r0
4.2 r < r2 and 0.3 < |m7-m81| < 0.5 and no other GLIMPSE source within r0
4.3 r < r2 and |m7-m81| < 0.3 and m82 > m7+1
                    (i.e. second GLIMPSE source fainter by atleast 1mag)

Flag = 3:  probable association
(separation > 2.3 && abs(m7-i4) < 0.5)||
(separation < 2.3 && abs(m7-i4) < 1.2 && abs(m7-i4)>0.5)||
(separation < 2.3 && m7 > 20)
3.1  r1 < r < r0 with |m7-m81| < 0.5 and no other GLIMPSE source within r0
3.2  r < r1 with 0.5 < |m7-m81| < 1.2 and no other GLIMPSE source within r0
3.3  r < r1 with 15micron detection only and no other GLIMPSE source within r0
3.4  r < r1 for second neighbour with |m7-m82| < 0.5 and |m7-m81| > 0.5
            - keep the second neighbour as association in this case

Flag = 2:  possible association
(separation < 3.8 && separation > 2.3 && abs(m7-i4) < 1.2 && abs(m7-i4)> 0.5)||
(separation < 2.3 && abs(m7-i4) > 1.2 && abs(m7-i4) < 2.2)||
(separation > 2.3 && m7 > 20)
2.1  r1 < r < r0 with 0.5 < |m7-m81| < 1.2 and no other GLIMPSE
2.2  r < r1 with 1.2 < |m7-m81| < 2.2 and no other GLIMPSE
2.3  r1 < r < r0 with 15micron only and no other GLIMPSE within r0
2.4  r1 < r < r0 for second neighbour with |m7-m82| < 1 and |m7-m81| > 1.5
            - keep the second neighbour as association in this case

Flag = 1:  hesitation to reject
(separation < 2.3 && abs(m7-i4) > 2.2 && abs(m7-i4) < 20)  ||
(separation > 2.3 && separation < 3.8 && abs(m7-i4) > 1.2) ||
(NULL_separation && m7 < 20 && m15 < 20)
1.1 r < r1 and |m7-m81| > 2.2
1.2 r1 < r < r2 and |m7-m81| > 1.2

- could be recovered with further info eg 24micron...
  e.g. 7 & 15 only and no GLIMPSE within r0
  or 7 with |m7-m81| > 2.2 and r < r0 and no other GLIMPSE

Flag = 0:  rejects - separate file
NULL_separation && (m7 > 20|| m15 > 20)


NOTES:
 - based on the GLIMPSE source quality flag: if the SQF bit 14 is set
  (i.e. 8192), then degrade the ultimate ISOGAL+GLIMPSE quality flag by 1.
 - if there is also a detection at LW3 then increase the flag value by 1
   - this is for sources with otherwise flags of 2 and 1 based on m7 values.
\end{verbatim}
\end{onecolumn}

\clearpage
\section{Multi-wavelength catalog for LN45 field}
\label{online_table}
\begin{table}[h]
\caption{An extract of the multi-wavelength catalog with the mass-loss rate, distance, extinction values and remarks on individual sources.  The complete catalog will be made available on the WWW at the CDS.
}
\label{tab_cat}
\begin{tabular}{|l|l|c|c|c|}
\hline
column name & units & source 1 & source 2 & source 3\\
\hline
id&number&2&14&187\\
\hline
R.A.&deg&217.5421&217.6252 &217.8831\\
Dec.&deg&-60.0549&-60.0626 &-60.113\\
\hline
ISOGAL-DENIS PSC1&name&PJ143010.1-600317&PJ143030.0-600345&PJ143131.9-600646\\
\hline
DENIS I&mag&12.35&99.99&13.96\\
DENIS J&mag&10.69 &99.99&8.62\\
DENIS $K_S$&mag&9.62&99.99&5.55\\
\hline
2MASS J&mag&10.718&11.55&8.565\\
2MASS H&mag&9.837&9.05&6.884\\
2MASS $K_S$&mag&9.626&7.878&6.035\\
\hline
IRAC1&mag&9.479&7.106&99.999\\
IRAC2&mag&9.6&7.062&99.999\\
IRAC3&mag&9.435&6.787&5.422\\
IRAC4&mag&9.39&6.724&5.277\\
\hline
LW2&mag&9.16&6.54&5.55\\
LW3&mag&99.99&5.78&4.38\\
\hline
MSX B1&Jy&&-1.412e1&-1.223e+1\\
MSX B2&Jy&&-7.623e+0&2.690e+0\\
MSX A&Jy&&9.725e-2&5.268e-1\\
MSX C&Jy&&-7.769e-1&4.335e-1\\
MSX D&Jy&&-5.570e-1&3.338e-1\\
MSX E&Jy&&-1.772e+0&4.387e-1\\
\hline
ISOGAL-GLIMPSE association&flag&5&4&5\\
source type&text&RGB&RGB&RGB\\
\hline
distance &pc&2369&6512&4000\\
extinction &mag $A_\mathrm{V}$&1.48&14.02&6.56\\
Absolute magnitude M$K_S$&mag&-2.37&-7.43&-7.55\\
log Mass-loss rate log($\dot{M}$)&log($M_\odot$/yr)&--&-7.69&-7.66\\
\hline
remarks&text&&AGB&IRAC1,2 saturated\\
\hline

\end{tabular}

\end{table}

\clearpage

\section{$J-K_S$ vs $J$ colour-magnitude diagrams selected by survey wavelengths}
\label{jjk_cmd_online}
\begin{figure*}[hb]
\includegraphics[width=0.73\textwidth]{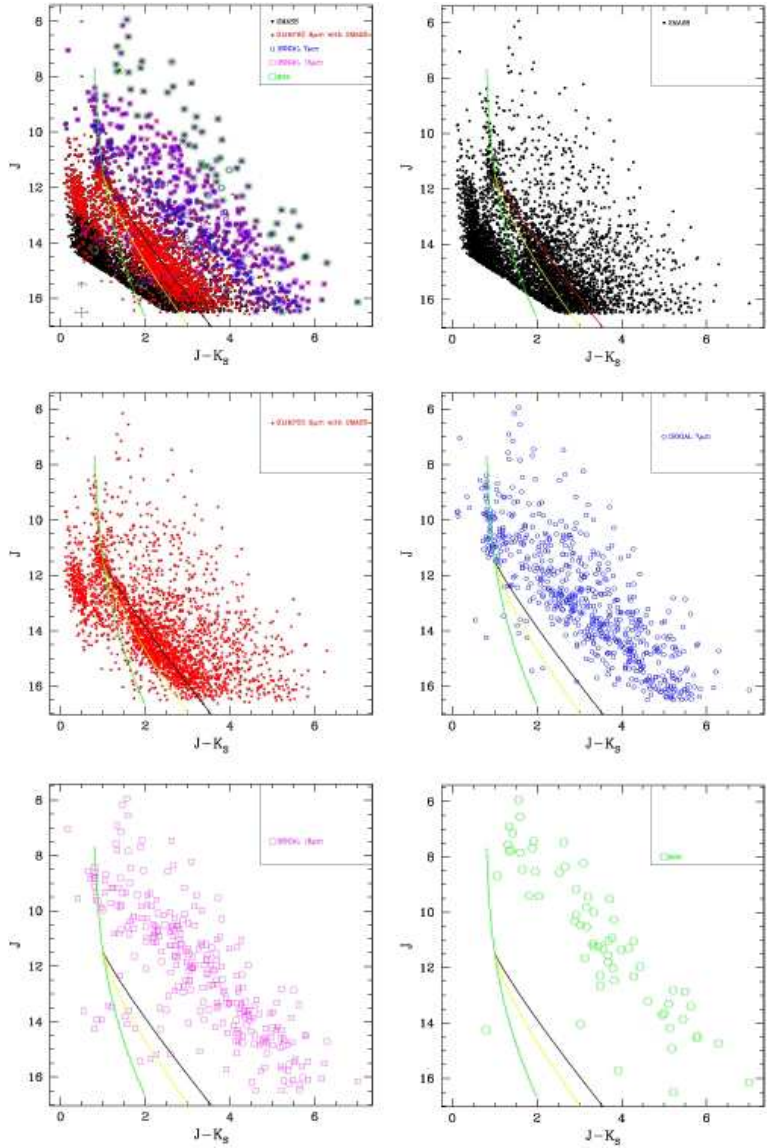}
\caption{$J-K_S$ vs $J$ colour-magnitude diagrams with the different survey detections displayed separately.  The solid lines are the red clump loci with standard $c_\mathrm{J}$ (i.e. local value of $c_\mathrm{J}$ being used throughout) displayed in green colour;  locus towards $b=0.5$ in yellow; locus towards $b=0.0$ in black.}
\label{fig_jk_j_surveys}
\end{figure*}

\clearpage
\section{Colour-colour diagrams to estimate mid-infrared extinction coefficients}
\begin{figure*}[hb]
\includegraphics[width=0.95\textwidth]{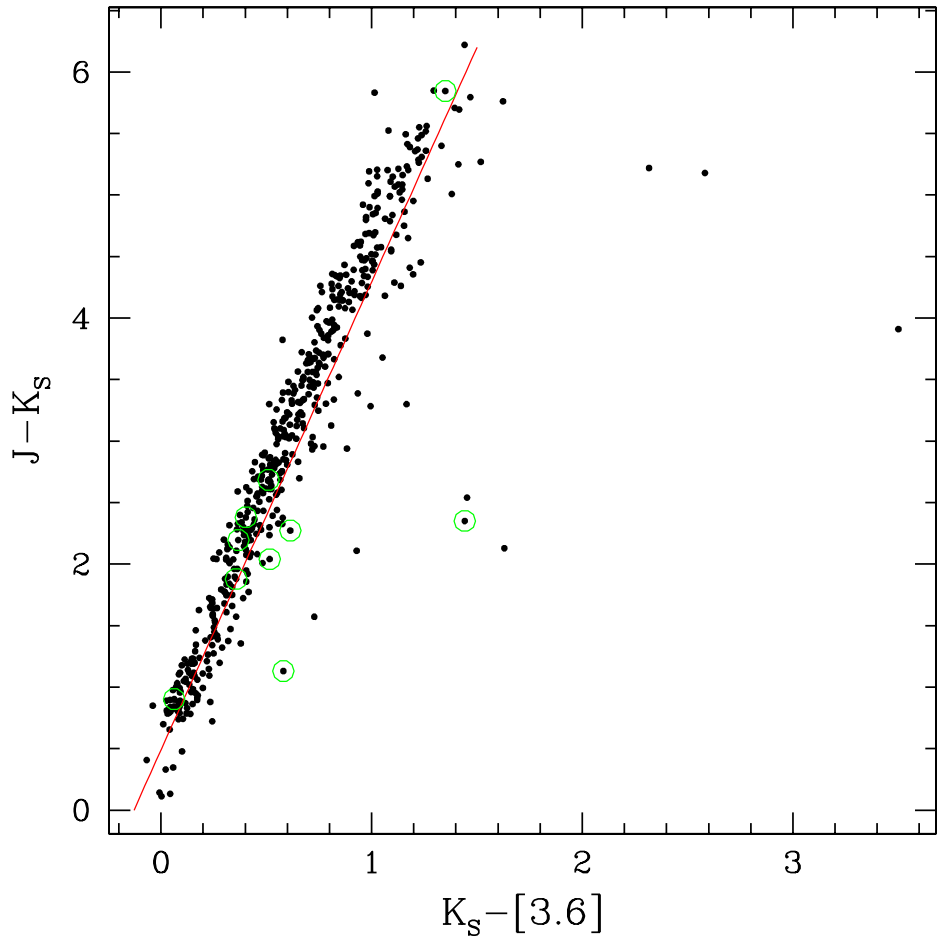}
\caption{$J-K_S$ vs $K_S-\mathrm{[3.6]}$ colour-colour diagram.  Sources detected only at [15] are marked with an additional green open circle.  The fit used to derive the extinction coefficient at 3.6$\mu$m is shown as a solid red line. }
\label{jk_ks3}
\end{figure*}

\begin{figure*}
\includegraphics[width=0.95\textwidth]{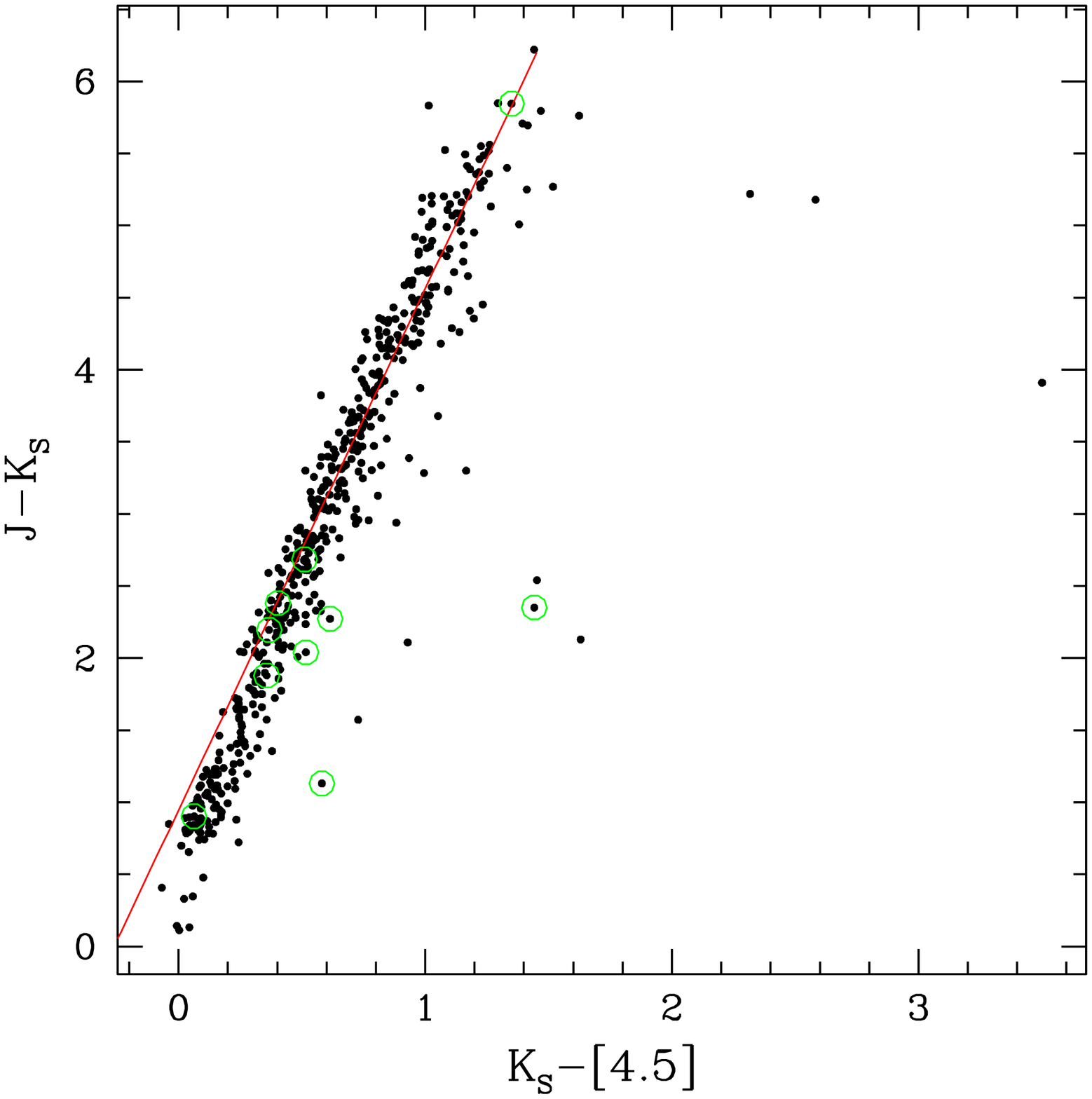}
\caption{$J-K_S$ vs $K_S-\mathrm{[4.5]}$ colour-colour diagram.  Sources detected only at [15] are marked with an additional green open circle.  The fit used to derive the extinction coefficient at 4.5$\mu$m is shown as a solid red line. }
\label{jk_ks4}
\end{figure*}

\begin{figure*}
\includegraphics[width=0.95\textwidth]{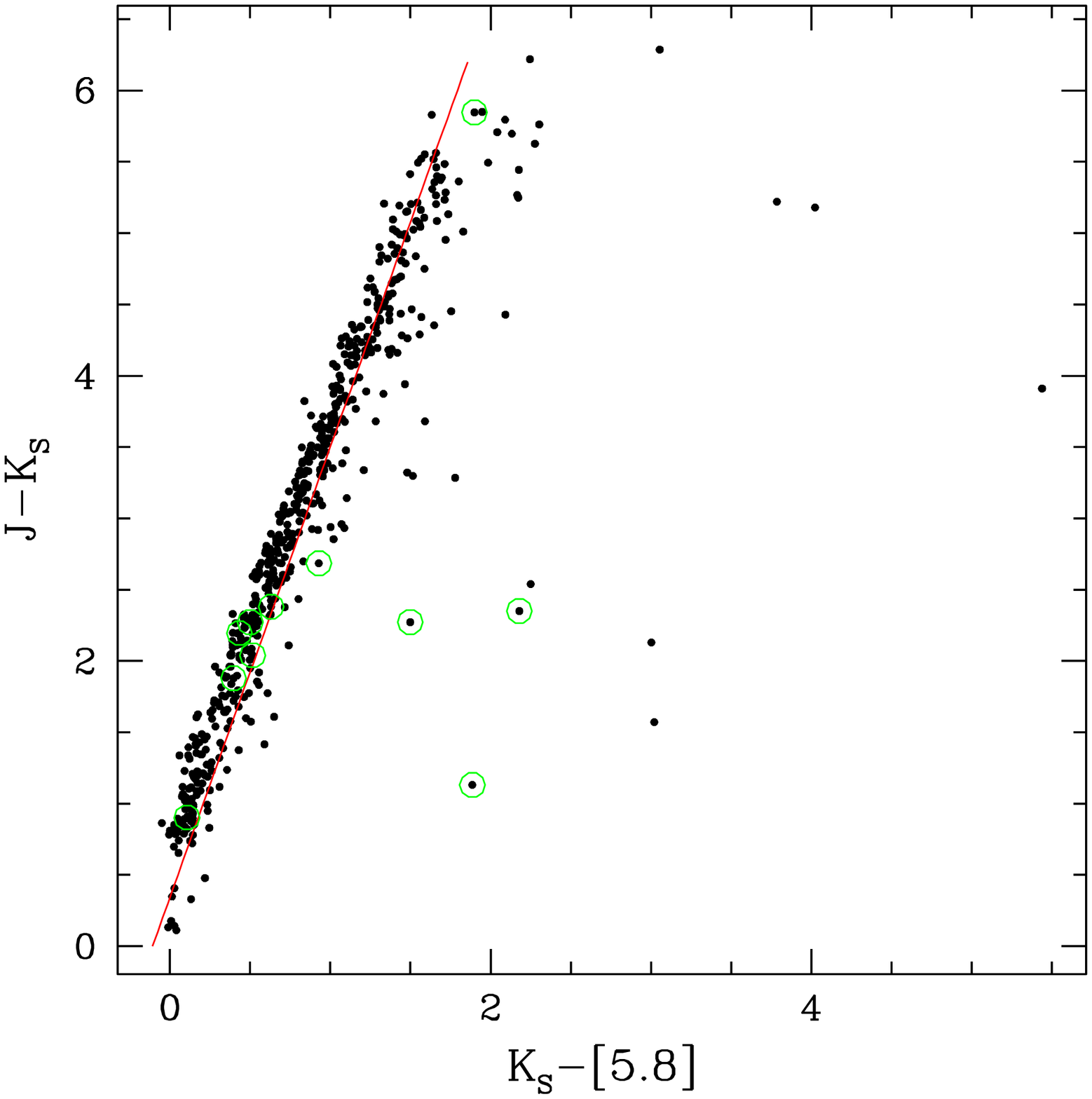}
\caption{$J-K_S$ vs $K_S-\mathrm{[5.8]}$ colour-colour diagram.  Sources detected only at [15] are marked with an additional green open circle.  The fit used to derive the extinction coefficient at 5.8$\mu$m is shown as a solid red line. }
\label{jk_ks5}
\end{figure*}

\clearpage
\section{$J-K_S$ vs $J$ colour-magnitude diagrams for various ranges in $b$}
\begin{figure*}[hb]
\includegraphics[width=0.99\textwidth]{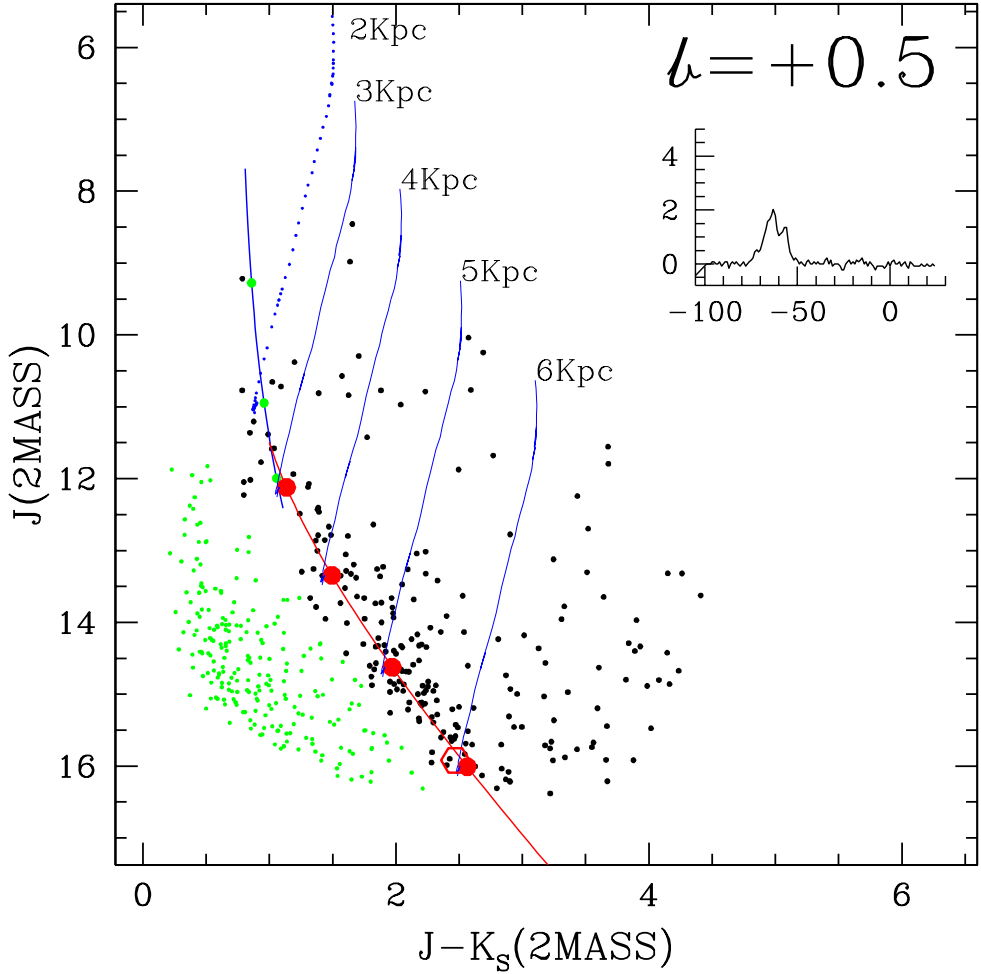}
\caption{$J-K_S$ vs $J$ CMD for the latitude ranges corresponding to the \element{C}\element{O} cell latitude limits ($b_{low}=0.4375$, $b_{high}=0.5$) centered at $b$=0.5. Also shown are the RGB/AGB isochrones at various distances  as labeled in the figure.  We use $A_\mathrm{V}=6.0(A_\mathrm{J}-A_\mathrm{K_S})$ (see Table \ref{tab_akl_ajk}).
We show with large open hexagons the points listed in table \ref{tab_av_dist_cmd_co}.  The corresponding \element{C}\element{O} spectrum is shown as an inset with the velocity scale (in km/sec) marked on the x axis. }
\label{fig_jk_j_CO05}
\end{figure*}
\begin{figure*}

\includegraphics[width=0.99\textwidth]{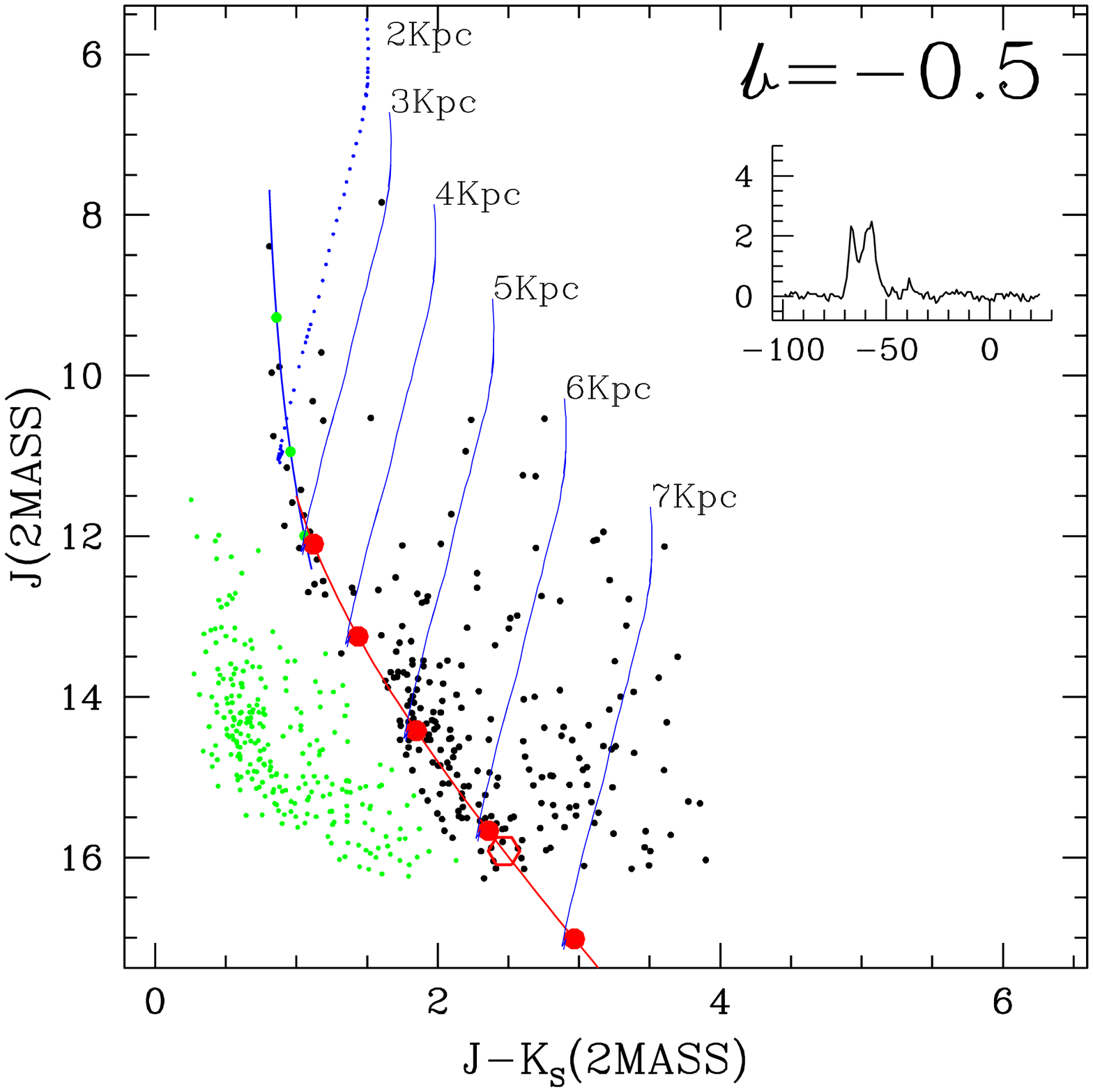}
\caption{$J-K_S$ vs $J$ CMD for the latitude ranges corresponding to the \element{C}\element{O} cell latitude limits ($b_{low}=-0.5$, $b_{high}=-0.4375$) centered at $b$=-0.5.  Rest of the details are as in Fig. \ref{fig_jk_j_CO05}}
\label{fig_jk_j_COm05}
\end{figure*}
\begin{figure*}
\includegraphics[width=0.99\textwidth]{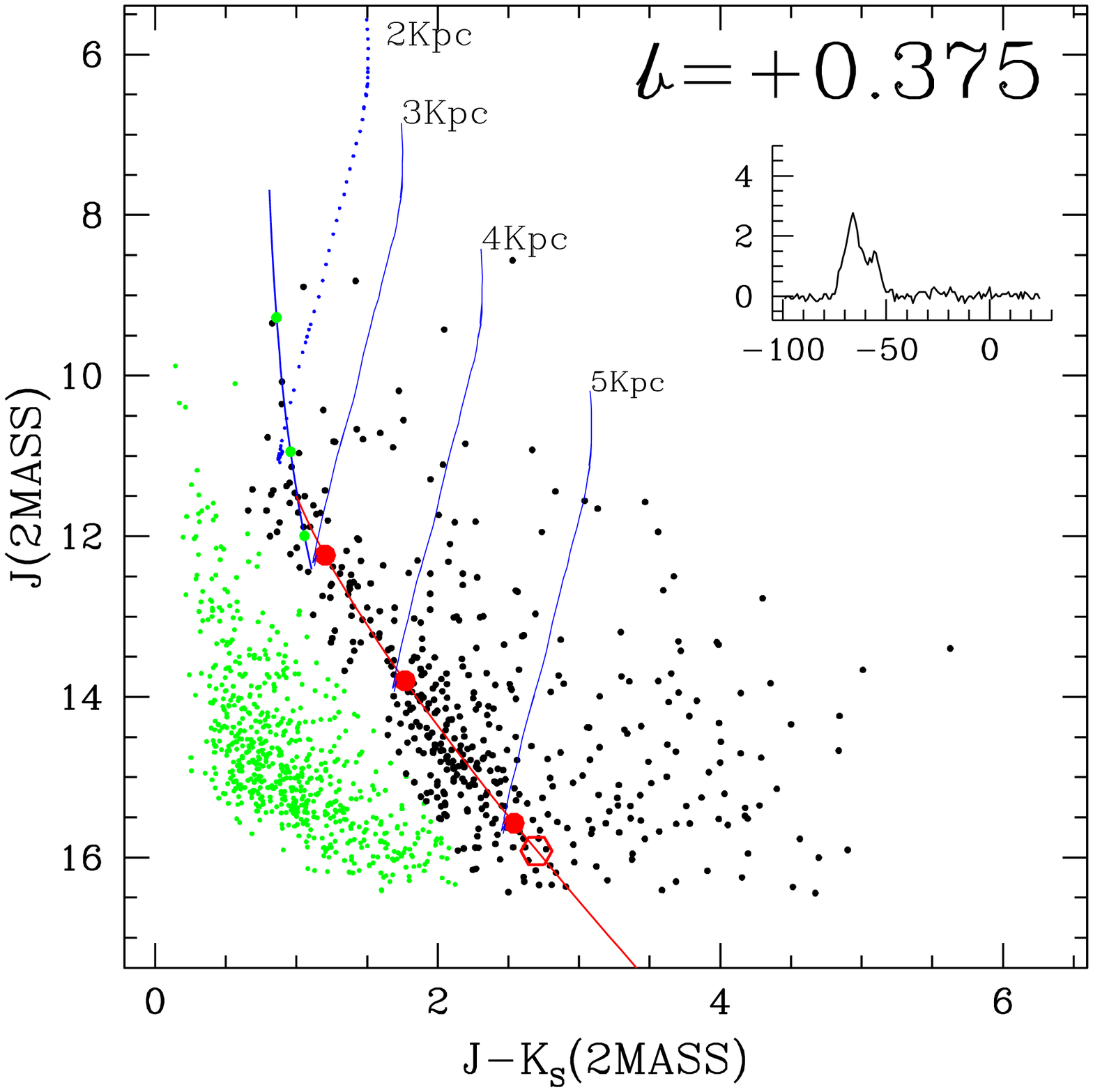}
\caption{$J-K_S$ vs $J$ CMD for the latitude ranges corresponding to the \element{C}\element{O} cell latitude limits ($b_{low}=0.3125$, $b_{high}=0.4375$) centered at $b$=0.375. Rest of the details are as in Fig. \ref{fig_jk_j_CO05}}
\label{fig_jk_j_CO0375}
\end{figure*}
\begin{figure*}
\includegraphics[width=0.99\textwidth]{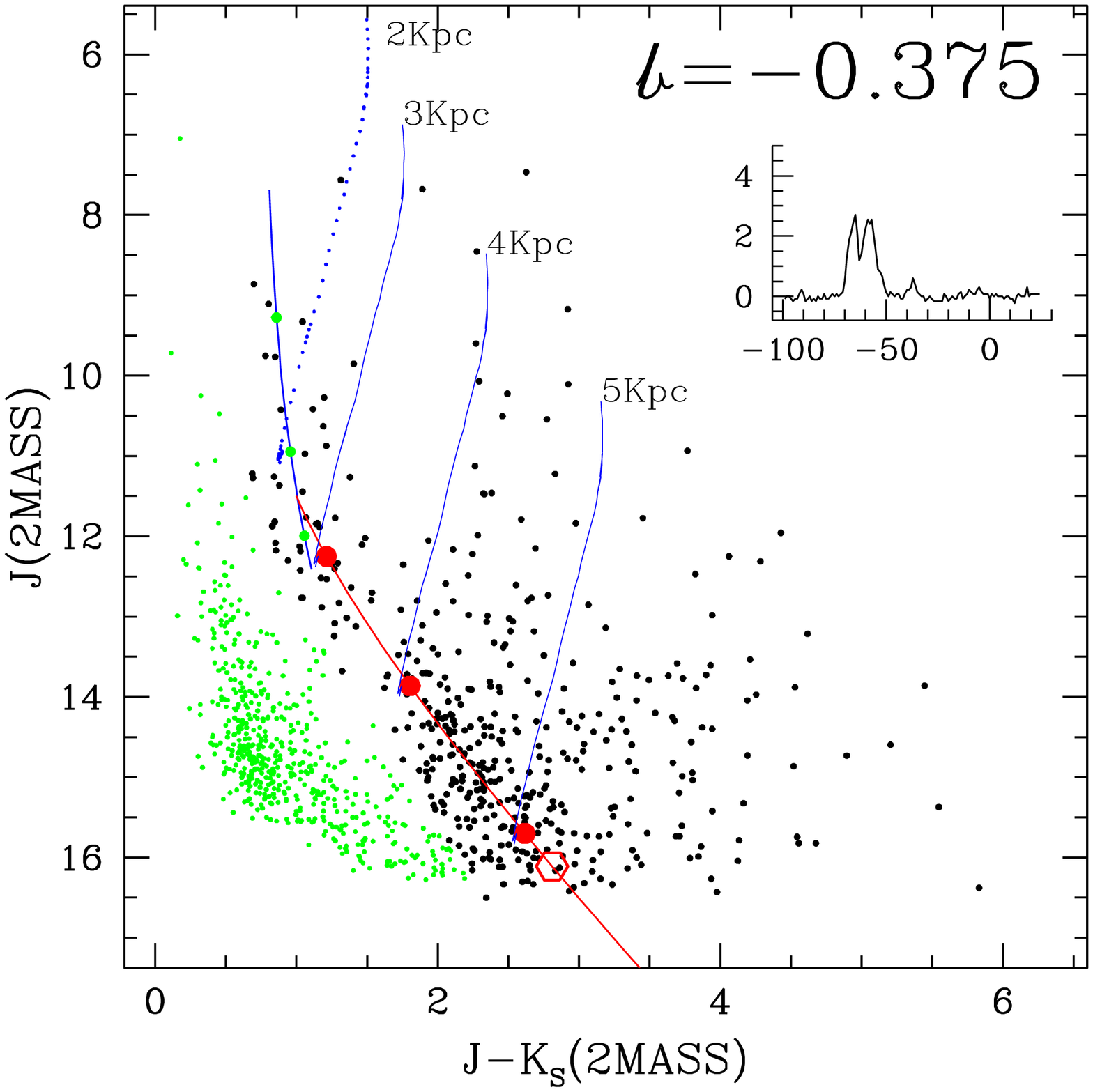}\\
\caption{$J-K_S$ vs $J$ CMD for the latitude ranges corresponding to the \element{C}\element{O} cell latitude limits ($b_{low}=-0.4375$, $b_{high}=-0.3125$) centered at $b$=-0.375. Rest of the details are as in Fig. \ref{fig_jk_j_CO05}}
\label{fig_jk_j_COm0375}
\end{figure*}
\begin{figure*}
\includegraphics[width=0.99\textwidth]{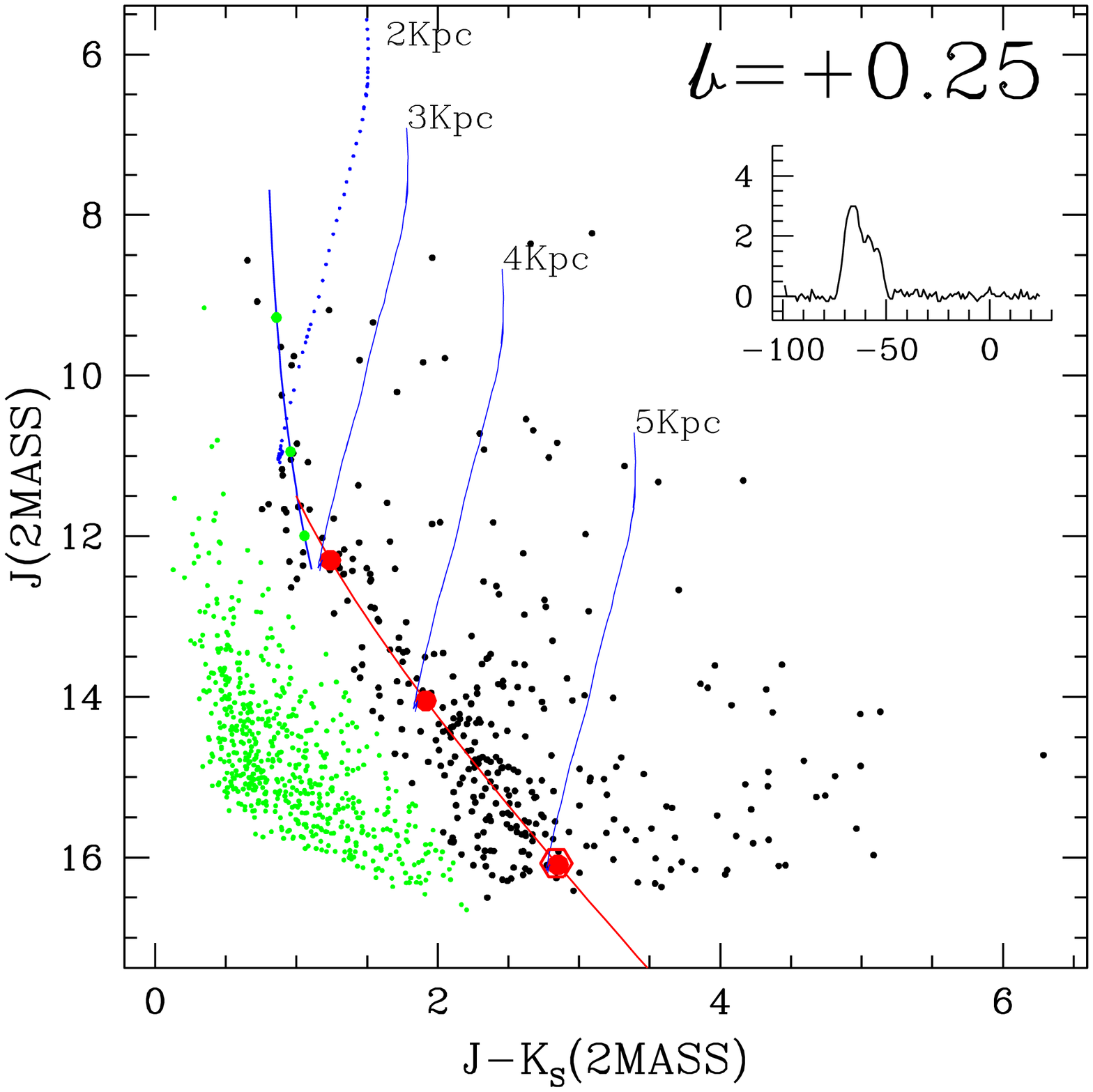}
\caption{$J-K_S$ vs $J$ CMD for the latitude ranges corresponding to the \element{C}\element{O} cell latitude limits ($b_{low}=0.1875$, $b_{high}=0.3125$) centered at $b$=0.25. Rest of the details are as in Fig. \ref{fig_jk_j_CO05}}
\label{fig_jk_j_CO025}
\end{figure*}
\begin{figure*}
\includegraphics[width=0.99\textwidth]{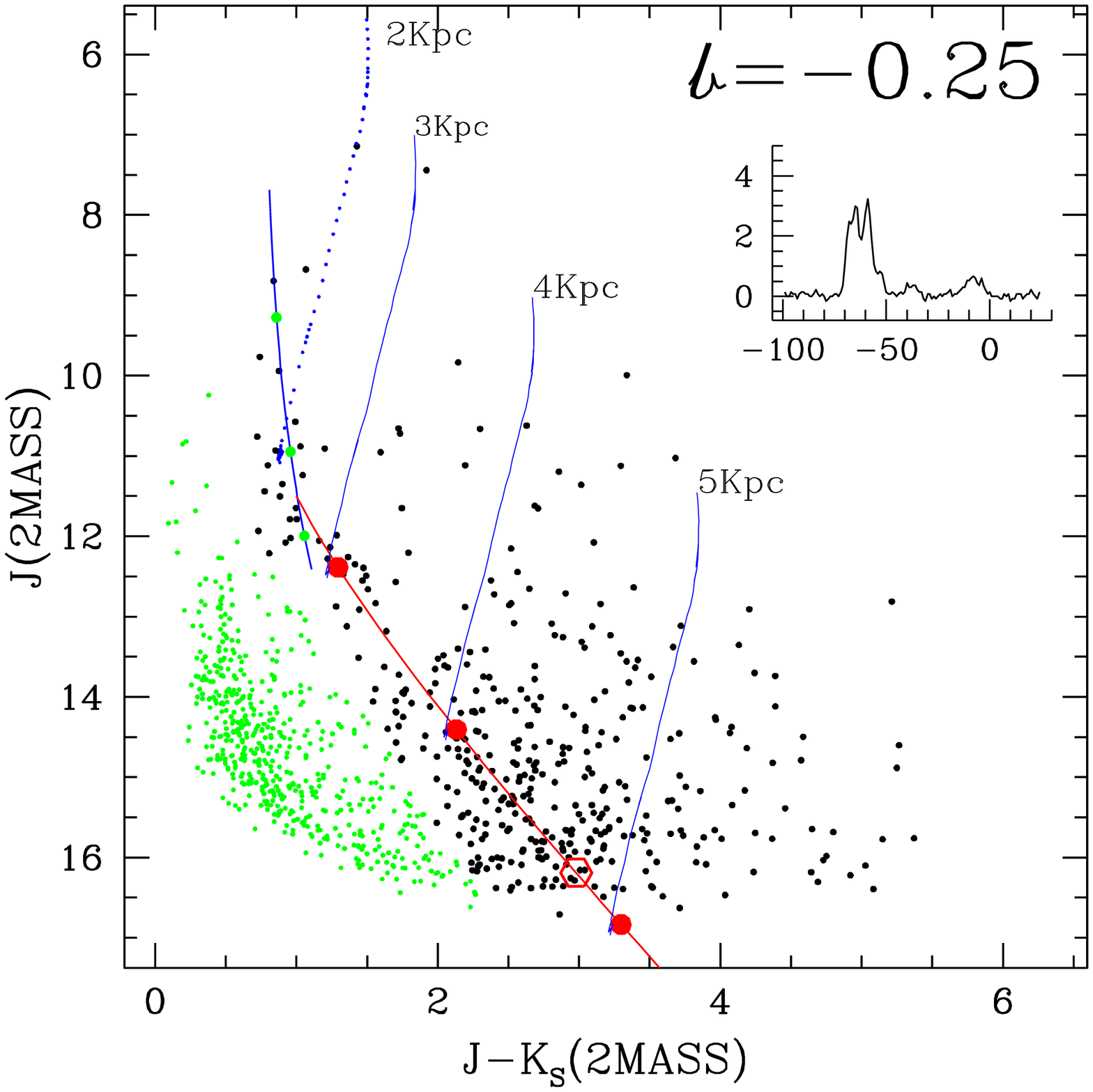}\\
\caption{ $J-K_S$ vs $J$ CMD for the latitude ranges corresponding to the \element{C}\element{O} cell latitude limits ($b_{low}=-0.3125$, $b_{high}=-0.1875$) centered at $b$=-0.25. Rest of the details are as in Fig. \ref{fig_jk_j_CO05}}
\label{fig_jk_j_COm025}
\end{figure*}
\begin{figure*}
\includegraphics[width=0.99\textwidth]{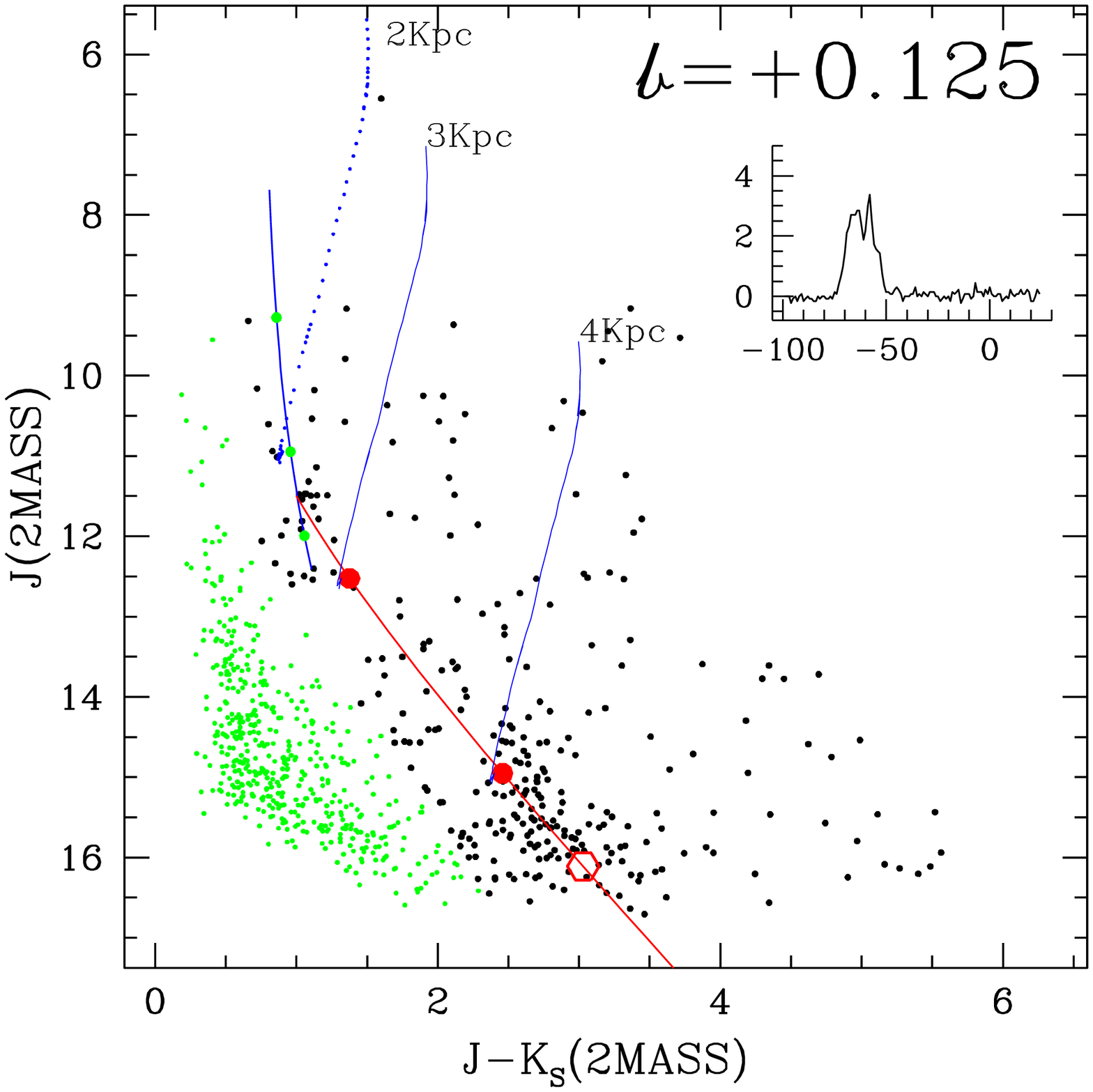}
\caption{ $J-K_S$ vs $J$ CMD for the latitude ranges corresponding to the \element{C}\element{O} cell latitude limits ($b_{low}=0.0625$, $b_{high}=0.1875$) centered at $b$=0.125. Rest of the details are as in Fig. \ref{fig_jk_j_CO05}}
\label{fig_jk_j_CO0125}
\end{figure*}
\begin{figure*}
\includegraphics[width=0.99\textwidth]{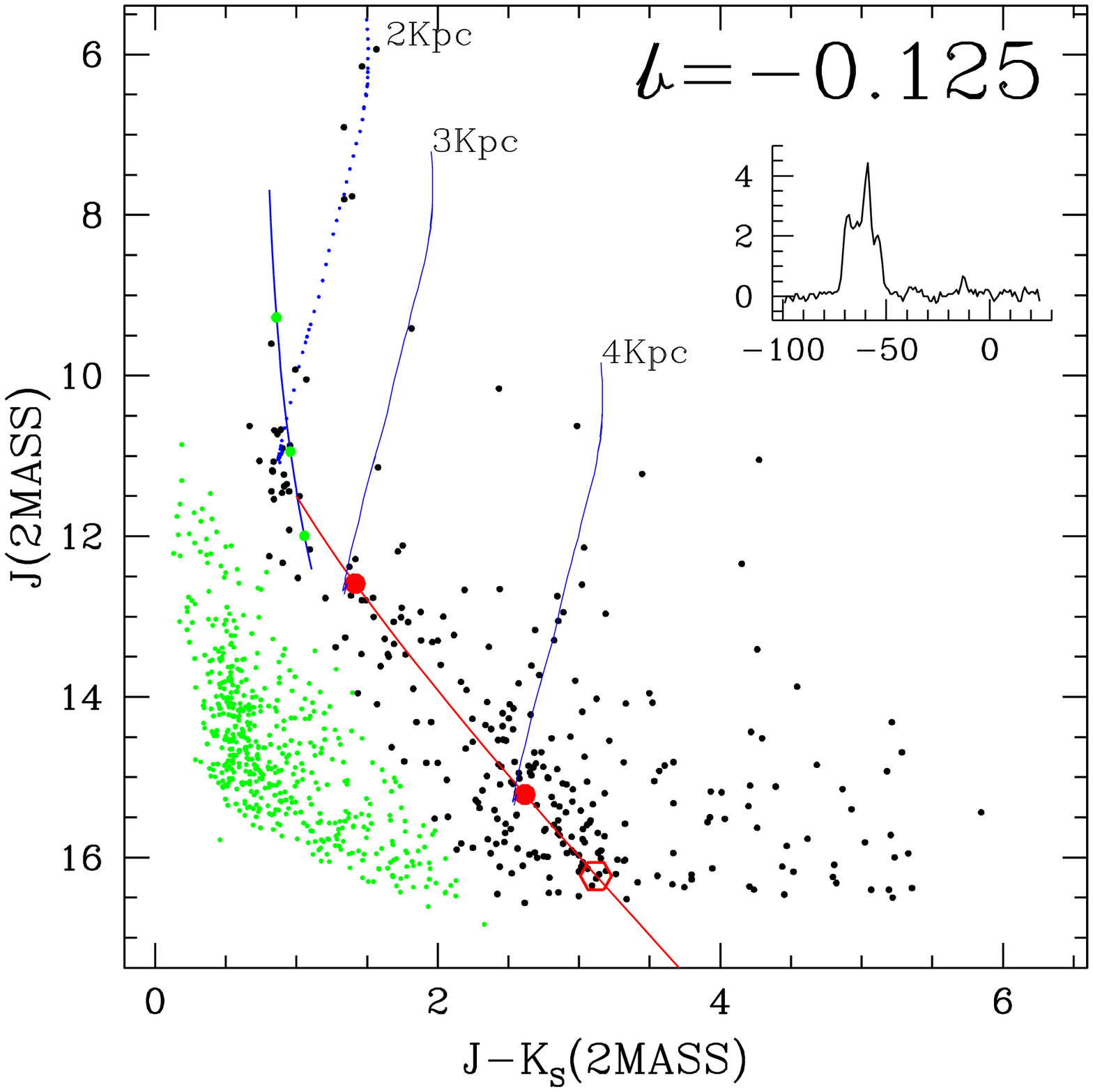}\\
\caption{$J-K_S$ vs $J$ CMD for the latitude ranges corresponding to the \element{C}\element{O} cell latitude limits ($b_{low}=-0.1875$, $b_{high}=-0.0625$) centered at $b$=-0.125. Rest of the details are as in Fig. \ref{fig_jk_j_CO05}}
\label{fig_jk_j_COm0125}
\end{figure*}
\begin{figure*}
\includegraphics[width=0.99\textwidth]{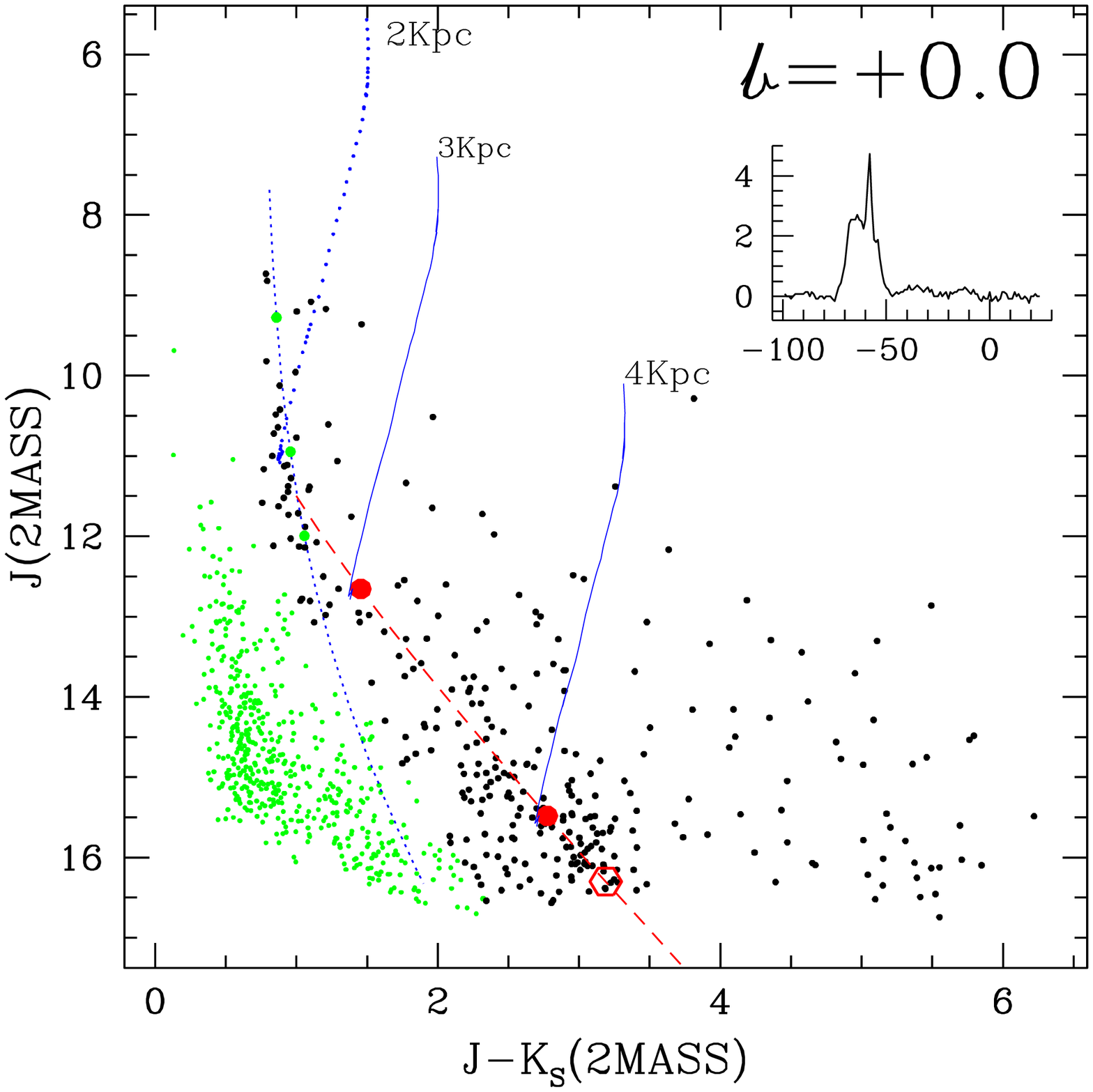}\\
\caption{$J-K_S$ vs $J$ CMD for the latitude ranges corresponding to the \element{C}\element{O} cell latitude limits ($b_{low}=-0.0625$, $b_{high}=0.0625$) centered at $b$=0.0. Rest of the details are as in Fig. \ref{fig_jk_j_CO05}}
\label{fig_jk_j_CO00}
\end{figure*}
\end{appendix}
\end{document}